\RequirePackage{lineno}
\documentclass[a4paper,10pt,twoside]{cpc-hepnp}
\usepackage{multicol}
\usepackage{graphicx}
\usepackage{booktabs}
\usepackage{amssymb}
\usepackage{bm}
\usepackage{color}
\usepackage{subfigure}
\usepackage{mathrsfs}
\usepackage{bbm}
\usepackage{amscd}
\usepackage[tbtags]{amsmath}
\usepackage{lastpage}

\begin{document}

\newcommand{\afs}{\alpha_s}
\newcommand{\bgp}{\beta\gamma}
\newcommand{\eff}{\varepsilon}
\newcommand{\sintht}{\sin{\theta}}
\newcommand{\costht}{\cos{\theta}}
\newcommand{\dedx}{dE/dx}

\newcommand{\probfc}{Prob_{\chi^2}}
\newcommand{\probpi}{Prob_{\pi}}
\newcommand{\probka}{Prob_{K}}
\newcommand{\probpr}{Prob_{p}}
\newcommand{\proball}{Prob_{all}}

\newcommand{\chicJ}{\chi_{cJ}}
\newcommand{\gchicJ}{\gamma\chi_{cJ}}
\newcommand{\gchica}{\gamma\chi_{c0}}
\newcommand{\gchicb}{\gamma\chi_{c1}}
\newcommand{\gchicc}{\gamma\chi_{c2}}
\newcommand{\hc}{h_c(^1p_1)}
\newcommand{\qqb}{q\bar{q}}
\newcommand{\uub}{u\bar{u}}
\newcommand{\ddb}{d\bar{d}}
\newcommand{\ssb}{\Sigma^0\bar{\Sigma}^0}
\newcommand{\ccb}{c\bar{c}}

\newcommand{\psipto}{\psi^{\prime}\rightarrow \pi^+\pi^- J/\psi}
\newcommand{\ptomm}{J/\psi\rightarrow \mu^+\mu^-}
\newcommand{\ppp}{\pi^+\pi^- \pi^0}
\newcommand{\pip}{\pi^+}
\newcommand{\pim}{\pi^-}
\newcommand{\kap}{K^+}
\newcommand{\kam}{K^-}
\newcommand{\ks}{K^0_s}
\newcommand{\pbar}{\bar{p}}
\newcommand{\jp}{J/\psi\rightarrow \gamma\pi^0}
\newcommand{\je}{J/\psi\rightarrow \gamma\eta}
\newcommand{\jep}{J/\psi\rightarrow \gamma\eta^{\prime}}

\newcommand{\LL}{\ell^+\ell^-}
\newcommand{\EE}{e^+e^-}
\newcommand{\MM}{\mu^+\mu^-}
\newcommand{\GG}{\gamma\gamma}
\newcommand{\TT}{\tau^+\tau^-}
\newcommand{\pp}{\pi^+\pi^-}
\newcommand{\kk}{K^+K^-}
\newcommand{\ppb}{p\bar{p}}
\newcommand{\gpp}{\gamma \pi^+\pi^-}
\newcommand{\gkk}{\gamma K^+K^-}
\newcommand{\gppb}{\gamma p\bar{p}}
\newcommand{\ggee}{\gamma\gamma e^+e^-}
\newcommand{\gguu}{\gamma\gamma\mu^+\mu^-}
\newcommand{\ggll}{\gamma\gamma l^+l^-}
\newcommand{\ppee}{\pi^+\pi^- e^+e^-}
\newcommand{\ppuu}{\pi^+\pi^-\mu^+\mu^-}
\newcommand{\etap}{\eta^{\prime}}
\newcommand{\gpi}{\gamma\pi^0}
\newcommand{\geta}{\gamma\eta}
\newcommand{\getap}{\gamma\etap}
\newcommand{\pppp}{\pi^+\pi^-\pi^+\pi^-}
\newcommand{\ppkk}{\pi^+\pi^-K^+K^-}
\newcommand{\pppr}{\pi^+\pi^-p\bar{p}}
\newcommand{\kkkk}{K^+K^-K^+K^-}
\newcommand{\kskp}{K^0_s K^+ \pi^- + c.c.}
\newcommand{\ppkp}{\pi^+\pi^-K^+ \pi^- + c.c.}
\newcommand{\ksks}{K^0_s K^0_s}
\newcommand{\dphi}{\phi\phi}
\newcommand{\phikk}{\phi K^+K^-}
\newcommand{\ppeta}{\pi^+\pi^-\eta}
\newcommand{\gpppp}{\gamma \pi^+\pi^-\pi^+\pi^-}
\newcommand{\gppkk}{\gamma \pi^+\pi^-K^+K^-}
\newcommand{\gpppr}{\gamma \pi^+\pi^-p\bar{p}}
\newcommand{\gkkkk}{\gamma K^+K^-K^+K^-}
\newcommand{\gkskp}{\gamma K^0_s K^+ \pi^- + c.c.}
\newcommand{\gppkp}{\gamma \pi^+\pi^-K^+ \pi^- + c.c.}
\newcommand{\gksks}{\gamma K^0_s K^0_s}
\newcommand{\gphiphi}{\gamma \phi\phi}

\newcommand{\tpp}{3(\pi^+\pi^-)}
\newcommand{\tppkk}{2(\pi^+\pi^-)(K^+K^-)}
\newcommand{\pptkk}{(\pi^+\pi^-)2(K^+K^-)}
\newcommand{\tkk}{3(K^+K^-)}
\newcommand{\gtpp}{\gamma 3(\pi^+\pi^-)}
\newcommand{\gtppkk}{\gamma 2(\pi^+\pi^-)(K^+K^-)}
\newcommand{\gpptkk}{\gamma (\pi^+\pi^-)2(K^+K^-)}
\newcommand{\gtkk}{\gamma 3(K^+K^-)}

\newcommand{\psp}{\psi(3686)}
\newcommand{\jpsi}{J/\psi}
\newcommand{\ar}{\rightarrow}
\newcommand{\lra}{\longrightarrow}
\newcommand{\jpsito}{J/\psi \rightarrow }
\newcommand{\ptoppjp}{J/\psi \rightarrow\pi^+\pi^- J/\psi}
\newcommand{\pspto}{\psi^\prime \rightarrow }
\newcommand{\ptop}{\psi'\rightarrow\pi^0 J/\psi}
\newcommand{\ptoeta}{\psi'\rightarrow\eta J/\psi}
\newcommand{\ecto}{\eta_c \rightarrow }
\newcommand{\ecpto}{\eta_c^\prime \rightarrow }
\newcommand{\xto}{X(3594) \rightarrow }
\newcommand{\chicJto}{\chi_{cJ} \rightarrow }
\newcommand{\chiczto}{\chi_{c0} \rightarrow }
\newcommand{\chicoto}{\chi_{c1} \rightarrow }
\newcommand{\chictto}{\chi_{c2} \rightarrow }
\newcommand{\pspp}{\psi^{\prime\prime}}
\newcommand{\ptochic}{\psi(2S)\ar \gamma\chi_{c1,2}}
\newcommand{\ppjpsi}{\pi^0\pi^0 J/\psi}
\newcommand{\utoeta}{\Upsilon^{\prime}\ar\eta\Upsilon}
\newcommand{\ww}{\omega\omega}
\newcommand{\wf}{\omega\phi}
\newcommand{\ff}{\phi\phi}
\newcommand{\npsp}{N_{\psp}}
\newcommand{\llb}{\Lambda\bar{\Lambda}}
\newcommand{\llbpi}{\llb\pi^0}
\newcommand{\llbeta}{\llb\eta}
\newcommand{\ppi}{p\pi^-}
\newcommand{\pbpi}{\bar{p}\pi^+}
\newcommand{\lamb}{\bar{\Lambda}}
\def\ctup#1{$^{\cite{#1}}$}
\newcommand{\bfg}{\begin{figure}}
\newcommand{\efg}{\end{figure}}
\newcommand{\bitm}{\begin{itemize}}
\newcommand{\eitm}{\end{itemize}}
\newcommand{\bnum}{\begin{enumerate}}
\newcommand{\enum}{\end{enumerate}}
\newcommand{\btbl}{\begin{table}}
\newcommand{\etbl}{\end{table}}
\newcommand{\btbu}{\begin{tabular}}
\newcommand{\etbu}{\end{tabular}}
\newcommand{\bcl}{\begin{center}}
\newcommand{\ecl}{\end{center}}
\newcommand{\bbt}{\bibitem}
\newcommand{\beq}{\begin{equation}}
\newcommand{\eeq}{\end{equation}}
\newcommand{\beqr}{\begin{eqnarray}}
\newcommand{\eeqr}{\end{eqnarray}}
\newcommand{\red}{\color{red}}
\newcommand{\blue}{\color{blue}}
\newcommand{\yellow}{\color{yellow}}
\newcommand{\green}{\color{green}}
\newcommand{\purple}{\color{purple}}
\newcommand{\brown}{\color{brown}}
\newcommand{\black}{\color{black}}

\title{\boldmath Determination of the number of $\psp$ events at BESIII}
\author
{
M.~Ablikim $^{1}$, M.~N.~Achasov$^{9,e}$, S. ~Ahmed$^{14}$, X.~C.~Ai $^{1}$,
O.~Albayrak$^{5}$, M.~Albrecht$^{4}$, \and D.~J.~Ambrose$^{44}$,
A.~Amoroso$^{49A,49C}$, F.~F.~An $^{1}$, Q.~An $^{46,a}$
% ...
}
\maketitle
\address{%
\vspace{0.2cm}
(BESIII Collaboration)\\
\vspace{0.2cm}
\it{
$^{1}$ Institute of High Energy Physics, Beijing 100049, People's Republic of China\\
$^{2}$ Beihang University, Beijing 100191, People's Republic of China\\
$^{3}$ Beijing Institute of Petrochemical Technology, Beijing 102617, People's Republic of China\\
$^{4}$ Bochum Ruhr-University, D-44780 Bochum, Germany\\
$^{5}$ Carnegie Mellon University, Pittsburgh, Pennsylvania 15213, USA\\
$^{6}$ Central China Normal University, Wuhan 430079, People's Republic of China\\
$^{7}$ China Center of Advanced Science and Technology, Beijing 100190, People's Republic of China\\
$^{8}$ COMSATS Institute of Information Technology, Lahore, Defence Road, Off Raiwind Road, 54000 Lahore, Pakistan\\
$^{9}$ G.I. Budker Institute of Nuclear Physics SB RAS (BINP), Novosibirsk 630090, Russia\\
$^{10}$ GSI Helmholtzcentre for Heavy Ion Research GmbH, D-64291 Darmstadt, Germany\\
$^{11}$ Guangxi Normal University, Guilin 541004, People's Republic of China\\
$^{12}$ Guangxi University, Nanning 530004, People's Republic of China\\
$^{13}$ Hangzhou Normal University, Hangzhou 310036, People's Republic of China\\
$^{14}$ Helmholtz Institute Mainz, Johann-Joachim-Becher-Weg 45, D-55099 Mainz, Germany\\
$^{15}$ Henan Normal University, Xinxiang 453007, People's Republic of China\\
$^{16}$ Henan University of Science and Technology, Luoyang 471003, People's Republic of China\\
$^{17}$ Huangshan College, Huangshan 245000, People's Republic of China\\
$^{18}$ Hunan University, Changsha 410082, People's Republic of China\\
$^{19}$ Indiana University, Bloomington, Indiana 47405, USA\\
$^{20}$ (A)INFN Laboratori Nazionali di Frascati, I-00044, Frascati, Italy; (B)INFN and University of Perugia, I-06100, Perugia, Italy\\
$^{21}$ (A)INFN Sezione di Ferrara, I-44122, Ferrara, Italy; (B)University of Ferrara, I-44122, Ferrara, Italy\\
$^{22}$ Johannes Gutenberg University of Mainz, Johann-Joachim-Becher-Weg 45, D-55099 Mainz, Germany\\
$^{23}$ Joint Institute for Nuclear Research, 141980 Dubna, Moscow region, Russia\\
$^{24}$ Justus-Liebig-Universitaet Giessen, II. Physikalisches Institut, Heinrich-Buff-Ring 16, D-35392 Giessen, Germany\\
$^{25}$ KVI-CART, University of Groningen, NL-9747 AA Groningen, The Netherlands\\
$^{26}$ Lanzhou University, Lanzhou 730000, People's Republic of China\\
$^{27}$ Liaoning University, Shenyang 110036, People's Republic of China\\
$^{28}$ Nanjing Normal University, Nanjing 210023, People's Republic of China\\
$^{29}$ Nanjing University, Nanjing 210093, People's Republic of China\\
$^{30}$ Nankai University, Tianjin 300071, People's Republic of China\\
$^{31}$ Peking University, Beijing 100871, People's Republic of China\\
$^{32}$ Seoul National University, Seoul, 151-747 Korea\\
$^{33}$ Shandong University, Jinan 250100, People's Republic of China\\
$^{34}$ Shanghai Jiao Tong University, Shanghai 200240, People's Republic of China\\
$^{35}$ Shanxi University, Taiyuan 030006, People's Republic of China\\
$^{36}$ Sichuan University, Chengdu 610064, People's Republic of China\\
$^{37}$ Soochow University, Suzhou 215006, People's Republic of China\\
$^{38}$ Sun Yat-Sen University, Guangzhou 510275, People's Republic of China\\
$^{39}$ Tsinghua University, Beijing 100084, People's Republic of China\\
$^{40}$ (A)Ankara University, 06100 Tandogan, Ankara, Turkey; (B)Istanbul Bilgi University, 34060 Eyup, Istanbul, Turkey; (C)Uludag University, 16059 Bursa, Turkey; (D)Near East University, Nicosia, North Cyprus, Mersin 10, Turkey\\
$^{41}$ University of Chinese Academy of Sciences, Beijing 100049, People's Republic of China\\
$^{42}$ University of Hawaii, Honolulu, Hawaii 96822, USA\\
$^{43}$ University of Minnesota, Minneapolis, Minnesota 55455, USA\\
$^{44}$ University of Rochester, Rochester, New York 14627, USA\\
$^{45}$ University of Science and Technology Liaoning, Anshan 114051, People's Republic of China\\
$^{46}$ University of Science and Technology of China, Hefei 230026, People's Republic of China\\
$^{47}$ University of South China, Hengyang 421001, People's Republic of China\\
$^{48}$ University of the Punjab, Lahore-54590, Pakistan\\
$^{49}$ (A)University of Turin, I-10125, Turin, Italy; (B)University of Eastern Piedmont, I-15121, Alessandria, Italy; (C)INFN, I-10125, Turin, Italy\\
$^{50}$ Uppsala University, Box 516, SE-75120 Uppsala, Sweden\\
$^{51}$ Wuhan University, Wuhan 430072, People's Republic of China\\
$^{52}$ Zhejiang University, Hangzhou 310027, People's Republic of China\\
$^{53}$ Zhengzhou University, Zhengzhou 450001, People's Republic of China\\
\vspace{0.2cm}
$^{a}$ Also at State Key Laboratory of Particle Detection and Electronics, Beijing 100049, Hefei 230026, People's Republic of China\\
$^{b}$ Also at Bogazici University, 34342 Istanbul, Turkey\\
$^{c}$ Also at the Moscow Institute of Physics and Technology, Moscow 141700, Russia\\
$^{d}$ Also at the Functional Electronics Laboratory, Tomsk State University, Tomsk, 634050, Russia\\
$^{e}$ Also at the Novosibirsk State University, Novosibirsk, 630090, Russia\\
$^{f}$ Also at the NRC "Kurchatov Institute", PNPI, 188300, Gatchina, Russia\\
$^{g}$ Also at University of Texas at Dallas, Richardson, Texas 75083, USA\\
$^{h}$ Also at Istanbul Arel University, 34295 Istanbul, Turkey\\
$^{i}$ Also at Goethe University Frankfurt, 60323 Frankfurt am Main, Germany\\
}}

\begin{abstract}
The numbers of  $\psi(3686)$ events accumulated by the BESIII detector for the two rounds of data taking during 2009 and 2012 are determined to be $(107.0\pm0.8)\times 10^6$ and $(341.1\pm 2.1)\times 10^6$, respectively, by counting inclusive hadronic events, where the uncertainty is dominated by systematics and the statistical uncertainty is negligible. The number of events for the sample taken in 2009 is consistent with that of the previous measurement. The total number of $\psp$ events for the two data-taking periods is $(448.1\pm2.9)\times10^6$.
\end{abstract}

\begin{keyword}
$\psp$, inclusive process, hadronic events, Bhabha process
\end{keyword}

\begin{pacs}
13.25.Gv, 13.66.Bc, 13.20.Gd
\end{pacs}
\begin{multicols}{2}

\section{Introduction}
\label{sec:introduction}
During the years 2009 and 2012, in two data-taking periods, the BESIII experiment has accumulated the world's largest $\psi(3686)$ data sample in electron-positron collisions, which provides an excellent place to precisely study the transition of $\psi(3686)$ and the subsequent charmonium state, $e.g.$ $\chicJ, h_c$, and $\eta_c$, from $\psi(3686)$ transitions, as well as to search for rare decays for physics beyond the standard model. The number of $\psp$ events, $N_{\psp}$, is a crucial
and important parameter. The precision of $\psp$ will directly affect the accuracy of these measurements.

In this paper, we present the determination of $\npsp$ with
inclusive $\psp$ hadronic decays, whose branching ratio is
known rather precisely, $(97.85\pm 0.13)$\%, in the Particle Data Group (PDG)~\cite{PDG}.
In the analysis, the QED background yield under the $\psp$ peak is evaluated by analyzing
the two sets of off-resonance data samples taken close by in time, $i.e.$ $\sqrt{s} = 3.65$~GeV
collected in 2009 with an integrated luminosity of about 44\, pb$^{-1}$
and four energy points ranging from 3.542 to 3.600 GeV collected
in 2012 for $\tau$-mass scan with a total integrated luminosity of about 23\, pb$^{-1}$~\cite{taumass}, ~respectively.
The strategy for the background estimation has been successfully used in our previous measurement of the number of $\psp$ events collected in 2009~\cite{psp09}, since the energies of the
$\psp$ and off-resonance data samples are close.

\section{BESIII detector and Monte Carlo simulation}
BEPCII is a double-ring $\EE$ collider that has reached a peak
luminosity of $1\times10^{33}~\rm{cm}^{-2}\rm{s}^{-1}$ at a
center-of-mass energy of 3.773 GeV. The cylindrical core of the BESIII detector
consists of a helium-based main drift chamber (MDC), a plastic
scintillator time-of-flight (TOF) system, and a CsI(Tl)
electromagnetic calorimeter (EMC), which are all enclosed in a
superconducting solenoid magnet with a field strength of 1.0~T (0.9~T in 2012). The
solenoid is supported by an octagonal flux-return yoke with
resistive plate counter modules interleaved with steel as a muon
identifier. The acceptance for charged particles and photons is 93\%
over the $4\pi$ stereo angle. The charged-particle momentum
resolution at 1 GeV/$c$ is 0.5\%, and the photon energy resolution
at 1 GeV is 2.5\% (5\%) in the barrel (end-caps) of the EMC. More
details about the apparatus can be found in Ref.~\cite{BESIII}. The
MDC encountered the Malter effect due to cathode aging during
$\psi(3686)$ data taking during 2012. This effect was suppressed by mixing
about 0.2\% water vapor into the MDC operating gas~\cite{mdc}, and can be well modeled by  Monte Carlo (MC) simulation. The other sub-detectors worked well during 2009 and
2012.

The BESIII detector is modeled with a MC simulation based on
\textsc{geant}4~\cite{geant4}. The $\psi(3686)$ produced in the electron-positron collision are modeled  with the generator \textsc{kkmc}~\cite{kkmc}, which include the beam energy spread according to the measurement of BEPCII and the effect of initial state radiation (ISR).
The known decay modes of $\psi(3686)$ are generated with \textsc{evtgen}~\cite{evtgen} according to the branching ratios in the PDG~\cite{PDG}, while the remaining unknown decays are simulated using the \textsc{lundcharm} model~\cite{lund}. The MC generated events
are mixed with randomly triggered events recorded in data taking to take into account the possible effects from beam-related backgrounds, cosmic rays, electronic noises and random firings of detector channels.

\section{Event selection}
The data collected at the $\psi(3686)$ peak includes several different process, \emph{i.e.},  $\psp$ decays to hadrons or lepton pairs
($\EE, ~\MM$, and $\tau^+\tau^-$), radiative return to the $\jpsi$,
and $\jpsi$ decay due to the extended tail of the $\jpsi$ line shape, and non-resonant (QED) processes, namely continuum background, including $\EE\ar \gamma^* \ar$ hadrons,
lepton pairs, and $\EE\ar\EE +X$ ($X$=hadrons, lepton pairs).
The data also contains non-collison events, $e.g.$ cosmic rays, beam-associated
backgrounds, and electronic noises. The process of interest in this analysis is $\psp$ decaying into hadrons.

Charged tracks are required to be within 1 cm of the beam line in
the plane perpendicular to the beam and within $\pm10$ cm from the
Interaction Point (IP) in the beam direction. Showers reconstructed
in the EMC barrel region ($|\cos\theta|<0.80$) must have a minimum
energy of 25 MeV, while those in the end-caps
($0.86<|\cos\theta|<0.92$) must have at least 50 MeV. The photons in
the polar range between the barrel and end-caps are excluded due to the poor resolution. A requirement of the EMC
cluster timing [0,~700] ns is applied to suppress electronic noise
and energy deposits unrelated to the event.

At least one charged track is required for each candidate event. In the following, the selected events are classed into to three categories according to the multiplicity of charged tracks $N_i\text{good}$, $i.e.$, $N_\text{good}=1,~N_\text{good}=2$, and $N_\text{good}>2$, and named  type-I, II, III, respectively.

For type-III events, no further selection criteria is required.

For type-II events, the momentum of each track
is required to be less than 1.7~GeV/$c$ and the opening angle between
the two charged tracks is required to be less than $176^{\circ}$ to suppress
Bhabha and dimuon backgrounds.
Figures~\ref{pvspbb} and ~\ref{angbb} show the scatter plots of the momenta of the first charged track versus that of the second charged tracks, and the distribution of opening angle between the two charged
tracks for the type-II candidates from simulated Bhabha (top) and inclusive
$\psi(3686)$ (bottom) MC events, respectively.
Furthermore, a scaled energy requirement $E_{\rm visible}/E_{\rm
cm}>0.4$ is applied to suppress the low energy background (LEB),
comprised mostly of $\EE\ar\EE +X$ and double ISR events ($\EE \to
\gamma_{\rm ISR} \gamma_{\rm ISR} X$). Here, $E_{\rm visible}$
denotes the visible energy which is defined as the total energy of
all charged tracks (calculated with the track momentum by assuming
to be a pion) and neutral showers. $E_{\rm cm}$ denotes the
center-of-mass energy. Figure~\ref{evisdt}~(top) shows the $E_{\rm
visible}/E_{\rm cm}$ distributions of the type-II events for the
$\psp$ data and inclusive MC sample. The visible excess in
data at low energy is from the LEB events. Unless noted,
in all plots, the points with error bars denote the $\psp$ data collected in 2012
and the histogram denotes the corresponding MC simulation.

For type-I events, at least two additional
photons are required in the event. Compared to those events with
high multiplicity of charged tracks, the type-I sample suffers from more
backgrounds according to the vertex distribution of the charged
tracks. Thus, a neutral hadron $\pi^0$ candidate is required to suppress the background events~\cite{r09}, where the $\pi^0$ candidate is reconstructed by any $\gamma\gamma$ combination. In an event, only the one $\pi^0$ candidate, whose mass is closest to $\pi^0$ nominal value and satisfy $|M_{\GG}-M_{\pi^0}|<0.015$ MeV/c$^2$, is kept for further analysis. Figure~\ref{pi0} shows the $M_{\GG}$ distributions of selected $\pi^0$ candidate for the type-I events. With above selection criteria, the corresponding $E_{\rm
visible}/E_{\rm cm}$ distributions of the candidate events for the $\psi(3686)$ data
and inclusive MC sample are shown in Fig.~\ref{evisdt} (bottom). An additional requirement $E_{\rm visible}/E_{\rm cm}>$ 0.4 is applied to suppress the events from LEB.

\bcl
\includegraphics[width=5.0cm]{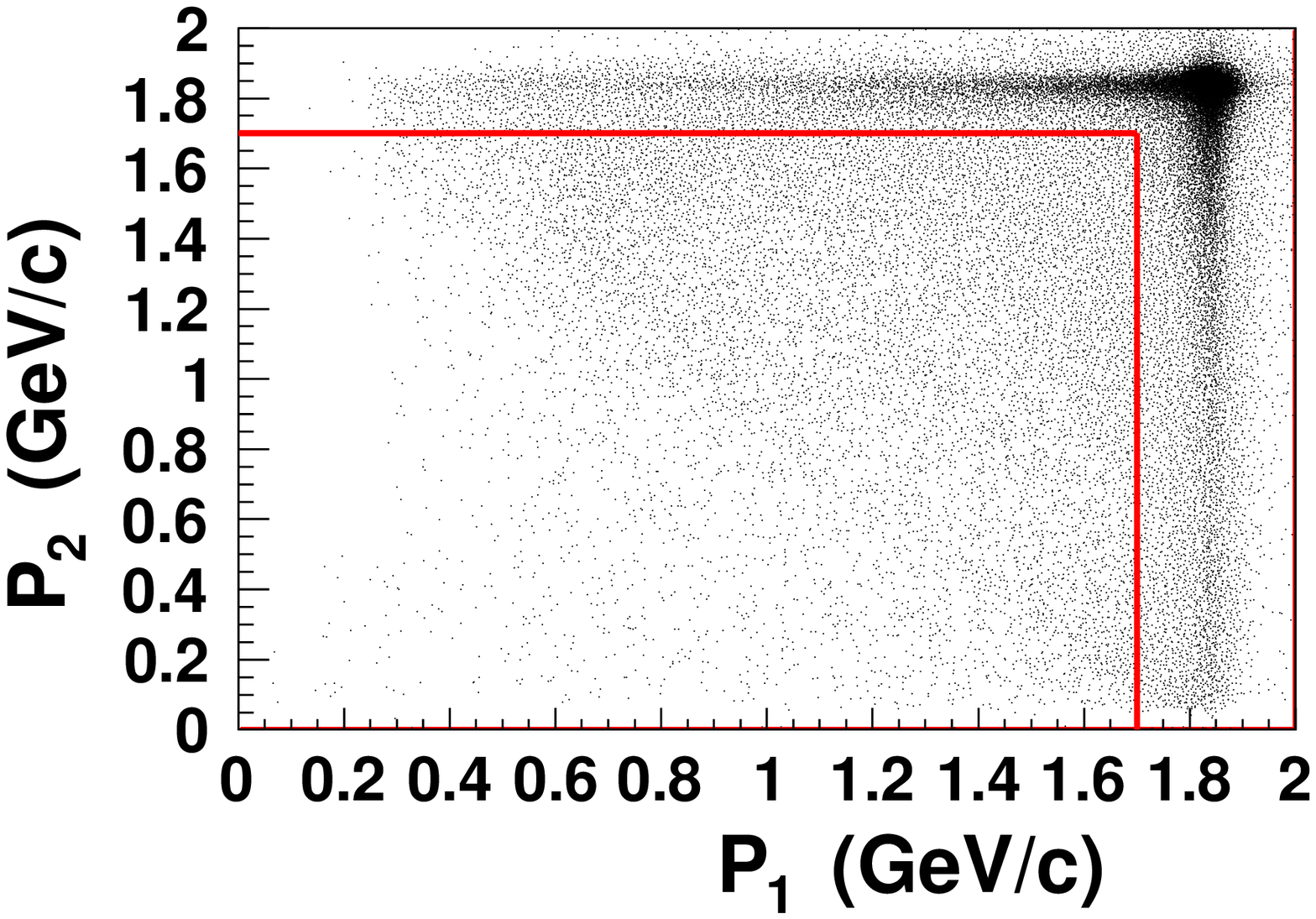}\hfill
\includegraphics[width=5.0cm]{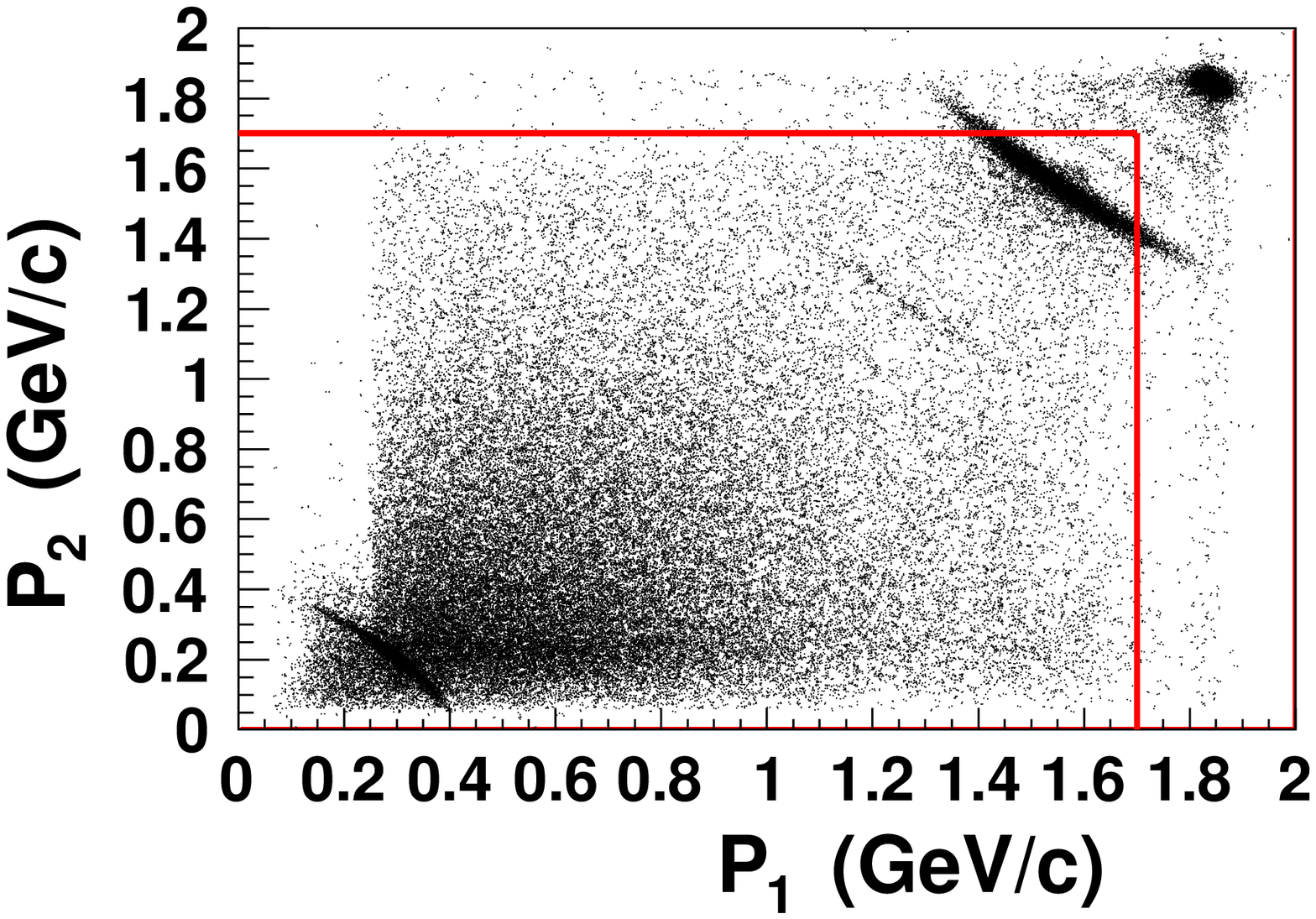}
\figcaption{\label{pvspbb}Scatter plots of the momenta of the first charged tracks versus that of second charged tracks of type-II candidates for Bhabha (top) and
inclusive $\psi(3686)$ (bottom) MC events. In the bottom plot, the event accumulation in the top-right corner comes from $\psp\to \EE, \MM$, while the different event bands nearby come from $\psp\to neutral +\jpsi, \jpsi\to \EE,\MM~ etc.$ The event band in the bottom-left comes from $\psp\to\ppjpsi,\jpsi\to \EE,\MM$ with lepton pairs missing. The horizontal and
vertical lines show the selection requirements to suppress Bhabha and
dimuon events.}
\ecl

\bcl
\includegraphics[width=6.0cm]{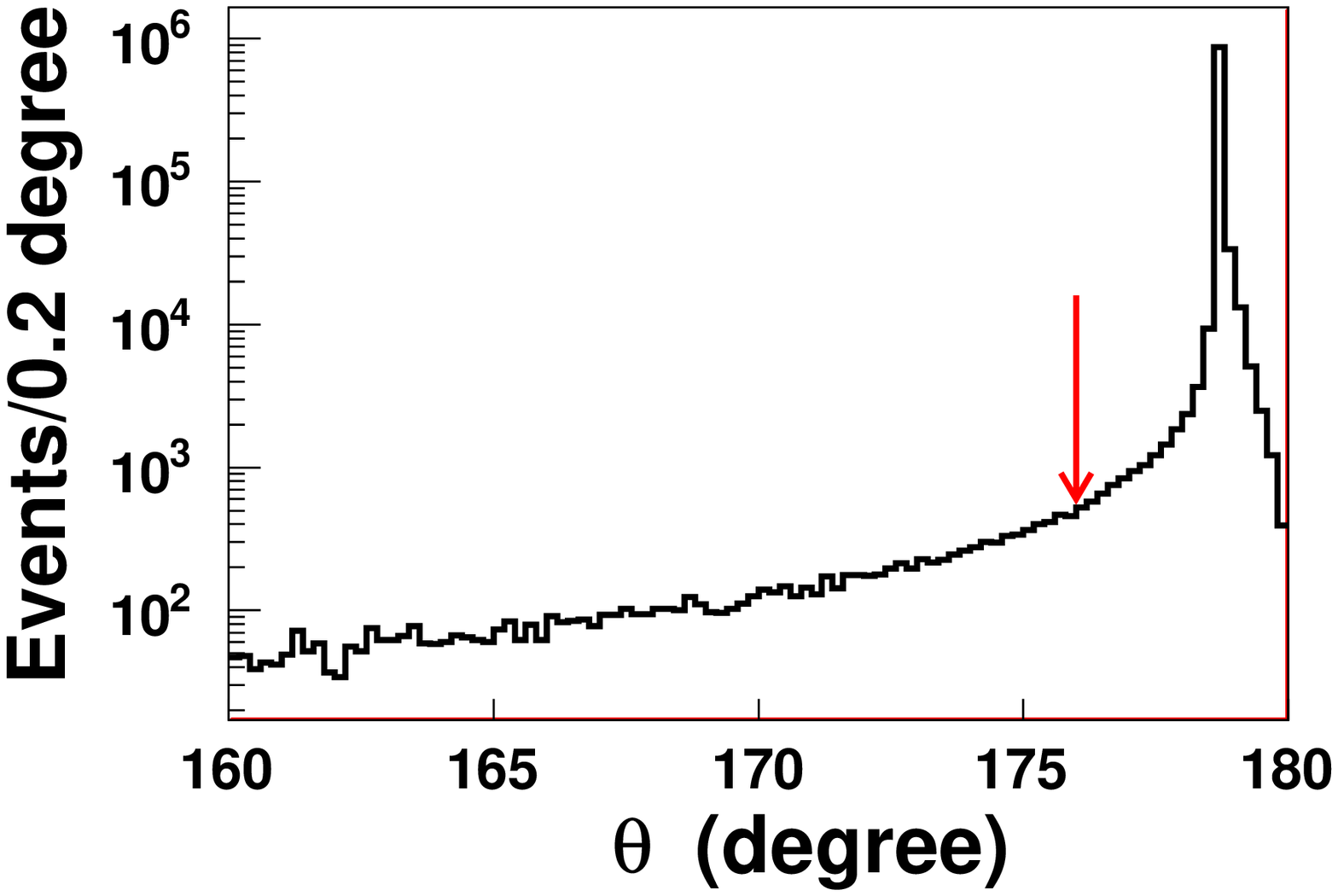}
\includegraphics[width=6.0cm]{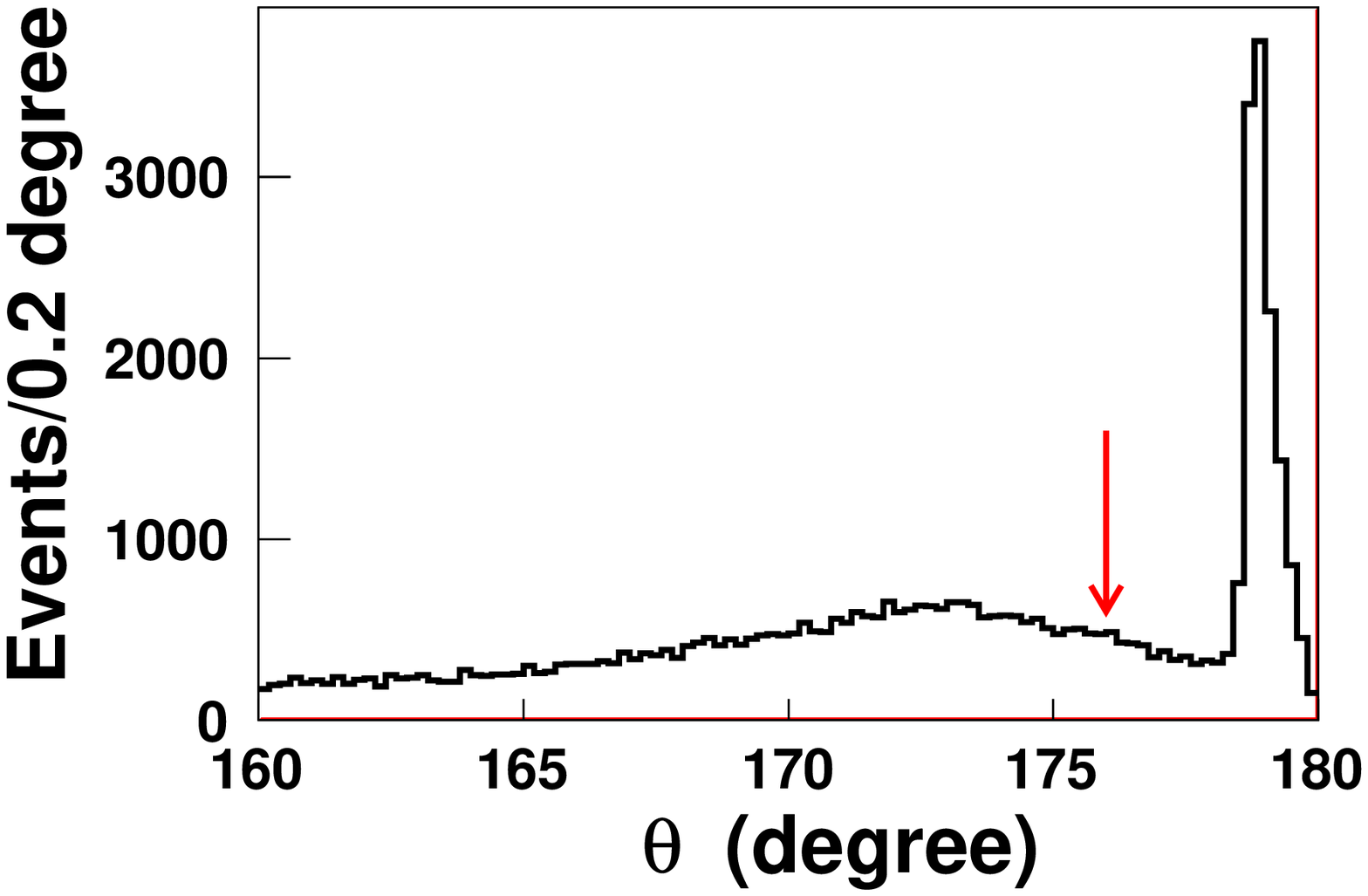}
\figcaption{\label{angbb}Distributions of the opening angle between
the two charged tracks for the type-II candidates from Bhabha (top) and inclusive
$\psi(3686)$ (bottom) MC events. The arrow shows the angle
requirement used to suppress Bhabha and dimuon events. }
\ecl
\bcl
\includegraphics[width=6.0cm]{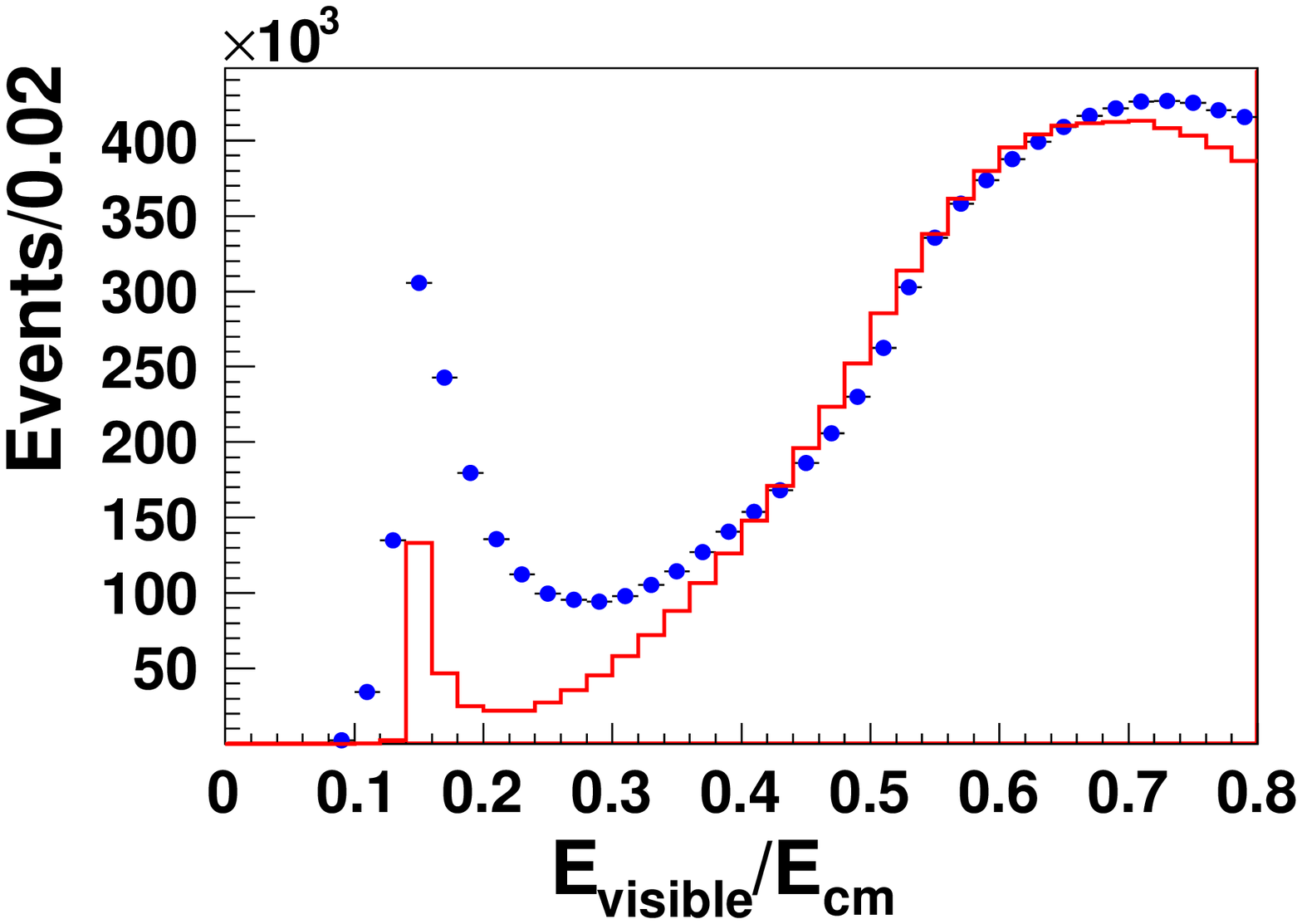}
\includegraphics[width=6.0cm]{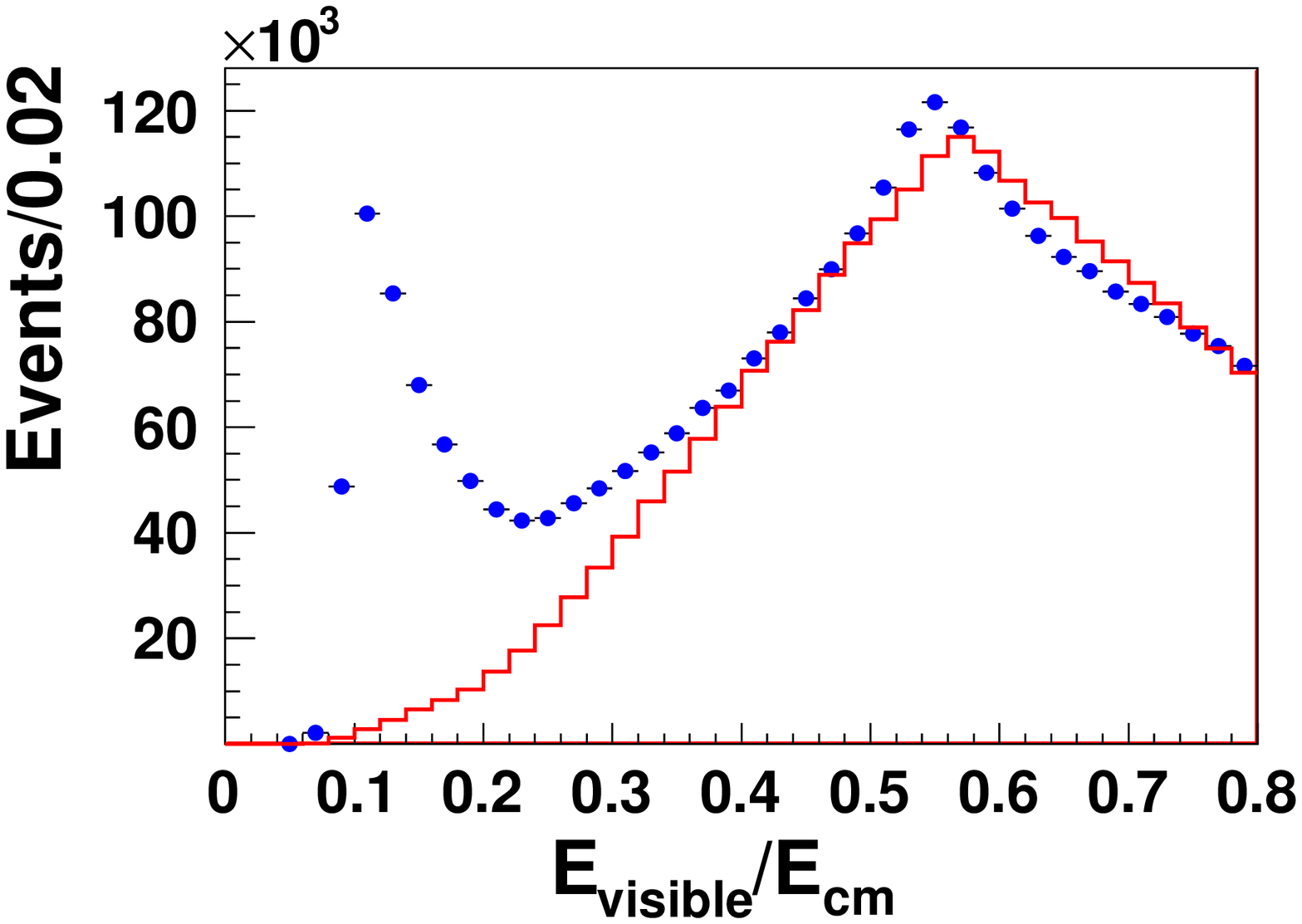}
\figcaption{\label{evisdt}Distribution of $E_{\rm visible}/E_{\rm cm}$
for the type-II (top) and type-I (bottom) events. The MC distributions are scaled arbitrarily to data with the same entries at
$E_{\rm visible}/E_{\rm cm}=0.4$.}
\ecl

\begin{center}
\includegraphics[width=6.0cm]{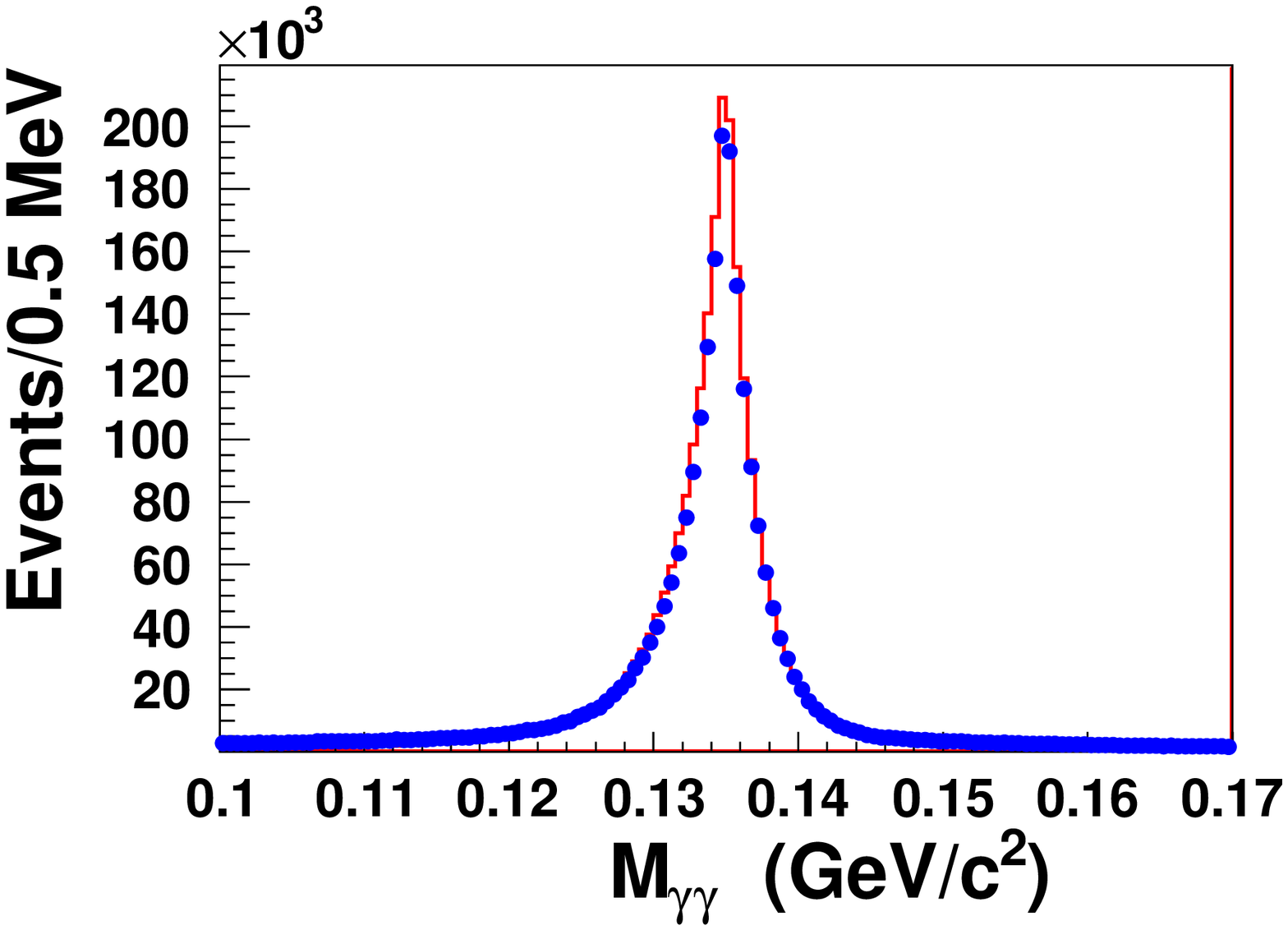}
\figcaption{Distribution of $M_{\GG}$ in the $\pi^0$ mass region
for the type-I events.}
\label{pi0}
\end{center}

To discriminate the non-collision background from the collision events, a variable,
the average vertex in Z direction is defined:
\[
\bar{V}_{Z}=\frac{\sum\limits^{N_{\rm good}}_{i=1}V_{Z}^i}{N_{\rm
good}},
\]
where $V_Z^i$ is the (signed) distance along the beam direction between the
point of closest approach of $i^{th}$ track and the IP. The $\bar{V}_{Z}$ distribution of the accepted hadronic events for the $\psi(3686)$ data is shown in the top plot of Fig.~\ref{z0}. The events satisfying
$|\bar{V}_Z|<4$~cm are taken as the signal, while the events in the
sideband region $6 <|\bar{V}_Z|<10$~cm are taken as non-collision
background events. The number of the observed hadronic events
($N^{\rm obs}$) is obtained by counting the events in the signal region ($N_{\rm signal}$) and subtract the non-collision background contribution estimated from the events in the sideband regions ($N_{\rm sideband}$).
\beq
\label{nhad} N^{\rm obs}=N_{\rm signal}-N_{\rm sideband}.
\eeq
We also try to determine the number of hadronic
events by fitting the $\bar{V}_Z$ distribution, where the signal event is described with a double Gaussian function, and the non-collision background is described with a second-order polynomial function. The resultant fit curves are shown in Fig.~\ref{z0}. This approach is used to be a cross check and to estimate the corresponding systematic uncertainty.

\bcl
\includegraphics[width=6.0cm]{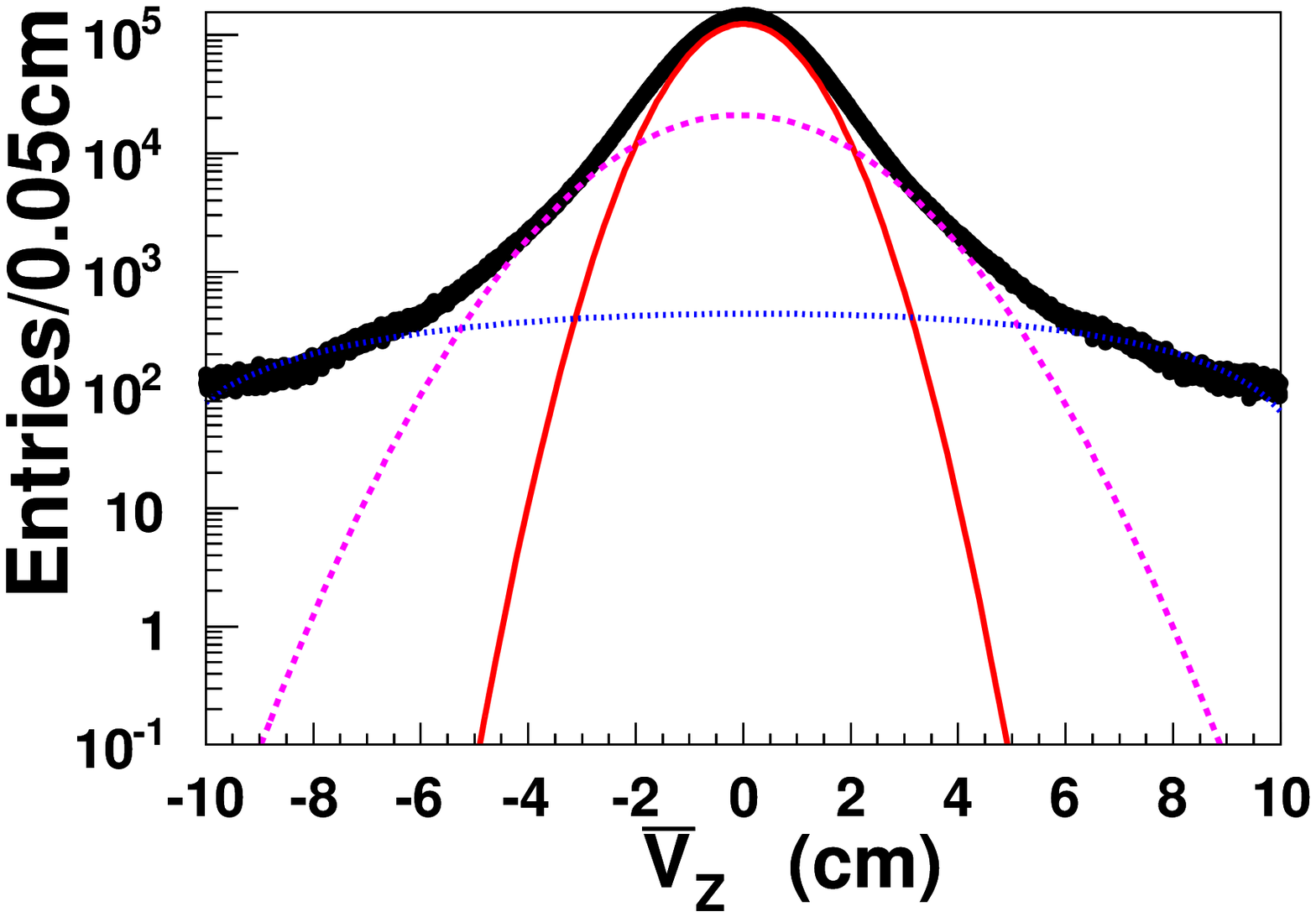}
\includegraphics[width=6.0cm]{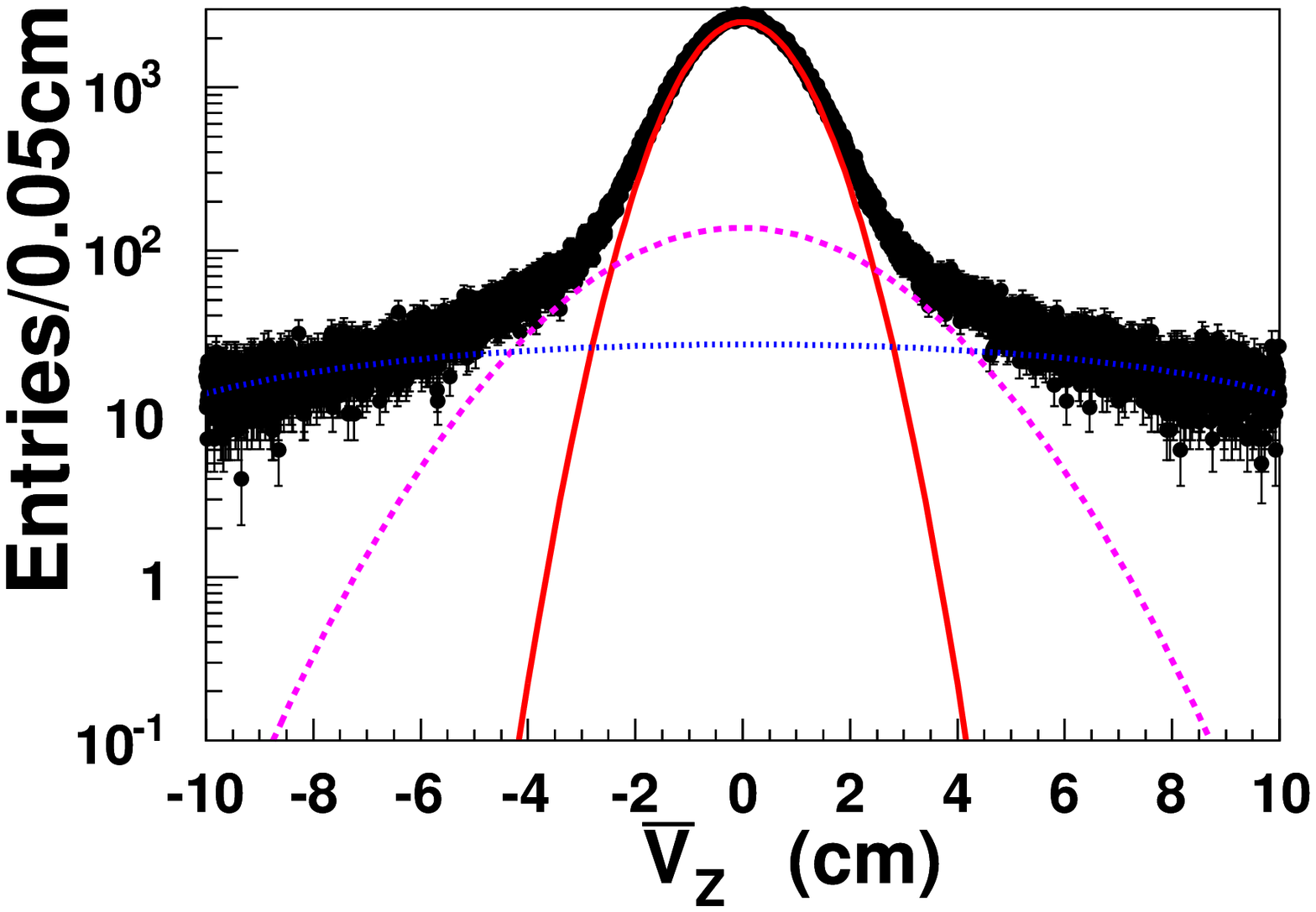}
\figcaption{Fits to the $\bar{V}_{Z}$ distributions of the
accepted hadronic events in the $\psp$ (top) and off-resonance
(bottom) data. The solid (red) and dashed (pink) curves show the double Gaussian line
shapes for the signal and the dotted (blue) lines show the
polynomial function for the non-collision events.}
\label{z0}
\ecl

\section{Background subtraction}

In general, the observed number of QED events can be estimated by \beq
\label{phbg} N^{\rm QED} = {\cal L}\cdot\sigma\cdot\epsilon, \eeq
where ${\cal L}$ is the integrated luminosity, $\sigma$ is the
theoretical cross section for the QED process, and $\epsilon$ is the
efficiency determined from a MC simulation. Alternatively, as mentioned in Section~{\ref{sec:introduction}}, the off-resonance data samples are used to estimate the continuum QED background yield.
We apply the same approaches to determine the yields of collision events and their uncertainty for the off-resonance data samples, which are dominant  from the continuum QED process.
With the above method, the effect of QED background is independent of the MC
simulation and the corresponding introduced systematic bias is expected to be small.

For the $\psi(3686)$ and off-resonance data samples, the
backgrounds from the radiative return to the $J/\psi$
and $\jpsi$ decay due to the extended tail are very similar due to the small difference in the center-of-mass
energies. The cross sections for this process are estimated to be
about 1.11 nb and 1.03 nb at the $\psp$ peak and the off-resonance
energy point, respectively. Detailed MC studies show that the
efficiencies for the known continuum processes are equal at these two energy
points. Thus, the off-resonance data sample are
used to estimate the number of both the continuum QED and $J/\psi$
decay backgrounds. Comparing to continuum QED processes, the fraction of background events from the radiative return
to the $J/\psi$ is very small, thus, a scaling factor, $f$, determined from the
integrated luminosity multiplied by a factor of $\frac{1}{s}$
($s=E_{\rm cm}^2$) is used to account for the energy dependence of
the cross section,
\beq\label{factor} f=\frac{{\cal L}_{\psp}}{{\cal
L}_{\rm off-resonance}}\cdot\frac{E_{\rm
off-resonance}^2}{E_{\psp}^2},
\eeq
where ${\cal L}_{\psp}$ and
${\cal L}_{\rm off-resonance}$ are the integrated luminosities for
the $\psp$ and off-resonance data samples, respectively, and
$E_{\psp}$ and $E_{\rm off-resonance}$ are the corresponding center-of-mass
energies. For the $\tau$-scan data, the average energy is determined to
be $\sqrt{s}=$3.572 GeV. The scaling factors $f$ are
determined to be 3.61 and 20.56 for the 2009 and 2012 data samples,
respectively. The slight variation of the cross section of the radiative return to the $\jpsi$ with center-of-mass energy
is negligible. The same is true for the background of the $\jpsi$ decay due to the extended tail.

The integrated luminosities of the data samples taken at different energy
points are determined from $\EE\ar\GG$ events using the following
selection criteria: Each event is required to have no good charged
track and at least two showers. The energies for the two most
energetic showers must be higher than 1.6 GeV and the cosine of the
polar angle of each electromagnetic shower must be within the region
$|\cos\theta|<0.8$. The two most energetic showers in the $\psp$
rest frame must be back to back, with azimuthal angles $||\phi_1 -
\phi_2 |-180^{\circ}| < 0.8^{\circ}$. The obtained luminosities are $161.63\pm 0.13\;\text{pb}^{-1}$ and $506.92\pm 0.23\;\text{pb}^{-1}$ for $\psp$ data taken during 2009 and 2012, respectively, while $43.88\pm 0.07\;\text{pb}^{-1}$ and $23.14\pm 0.05\;pb^{-1}$ for off-resonance data taken at $\sqrt{s}$=3.65 GeV and for $\tau$-scan data set, respectively. Here, the errors are statistical only. The systematic uncertainties
related to the luminosity  almost cancel in calculating
the scaling factor due to the small difference between the energy
points. The scaling factor can also be obtained using the
integrated luminosities determined with Bhabha events. The
difference in $f$ between these two methods is negligible.

In order to validate the LEB events remaining in the $\psp$ sample after applying the
$E_{\rm visible}/E_{\rm cm}$ selection, the LEB candidate events are
selected by requiring $E_{\rm visible}/E_{\rm cm}<0.35$, where
few QED events are expected. Figures~\ref{ebkg} (top) and (bottom)
show the comparisons of the $E_{\rm visible}/E_{\rm cm}$
distributions for the type-I (top) and type-II (bottom) LEB events
between the $\psi(3686)$ and the scaled off-resonance data samples taken in 2012.
The ratios of the event numbers between the $\psp$ peak and the off-resonance energy are 22.78 and 22.57 for the type I and type II events, respectively. Compared with the scaling factor obtained from the
integrated luminosity normalization in Eq.~(\ref{factor}), a difference of
about 10\% is found for the type-I and type-II events. Similar
differences are found for the 2009 data sample~\cite{psp09}.
Since the faction of LEB events in the selected sample is very small, the effect of this difference for the background estimation is very small and can be negligible.

The cross sections for $\EE\ar\TT$ are 0.67, 1.84, and 2.14~nb
at the $\tau$-scan energy ($\sqrt{s}=3.572\;\text{GeV}$ according to luminosity weighted average),
$\sqrt{s}=3.65\;\text{GeV}$ and the $\psp$ peak, respectively. Since the above energy points are closed to $\TT$ mass threshold, and the production cross sections does not follow an $1/s$ distribution. Thus, only a part of the $\EE\ar\TT$ background events have be considered by the off-resonance data samples. To compensate the background from the full background from $\EE\ar\TT$, we estimate its remaining contribution according the detection efficiency from the MC simulation and the cross section difference at off-resonance energy points and $\psi(3686)$ peak as well as the luminosity at $\psi(3686)$ peak. The estimated values are shown in Table~\ref{npsppart}, too.

The small number of the surviving events from $\psp\ar\EE,~\MM$, and
$\TT$ in data does not need to be explicitly subtracted since these
leptonic $\psp$ decays have been included in the inclusive MC samples, and their effects are considered in the detection efficiency.

\bcl
\includegraphics[width=6.0cm]{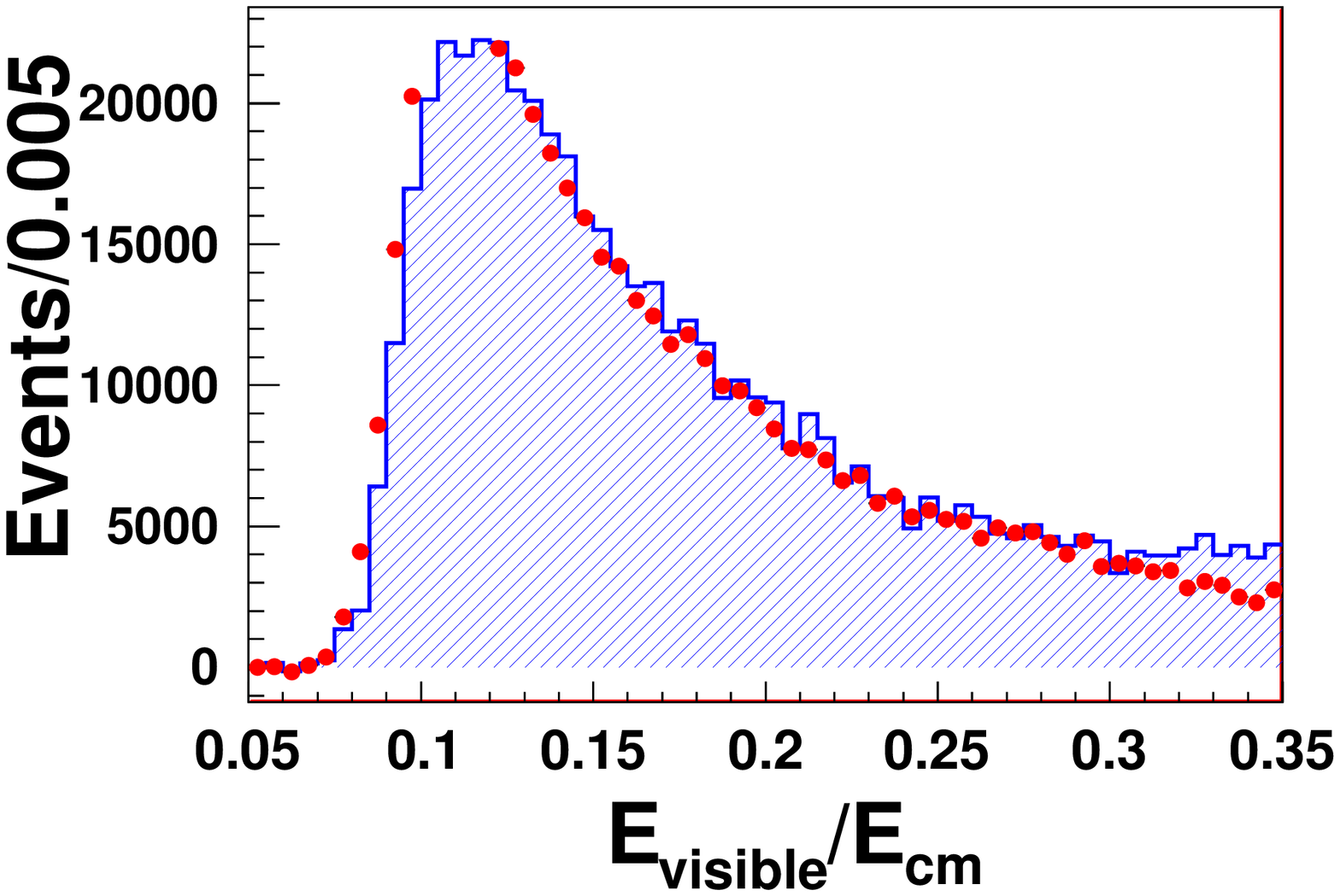}
\includegraphics[width=6.0cm]{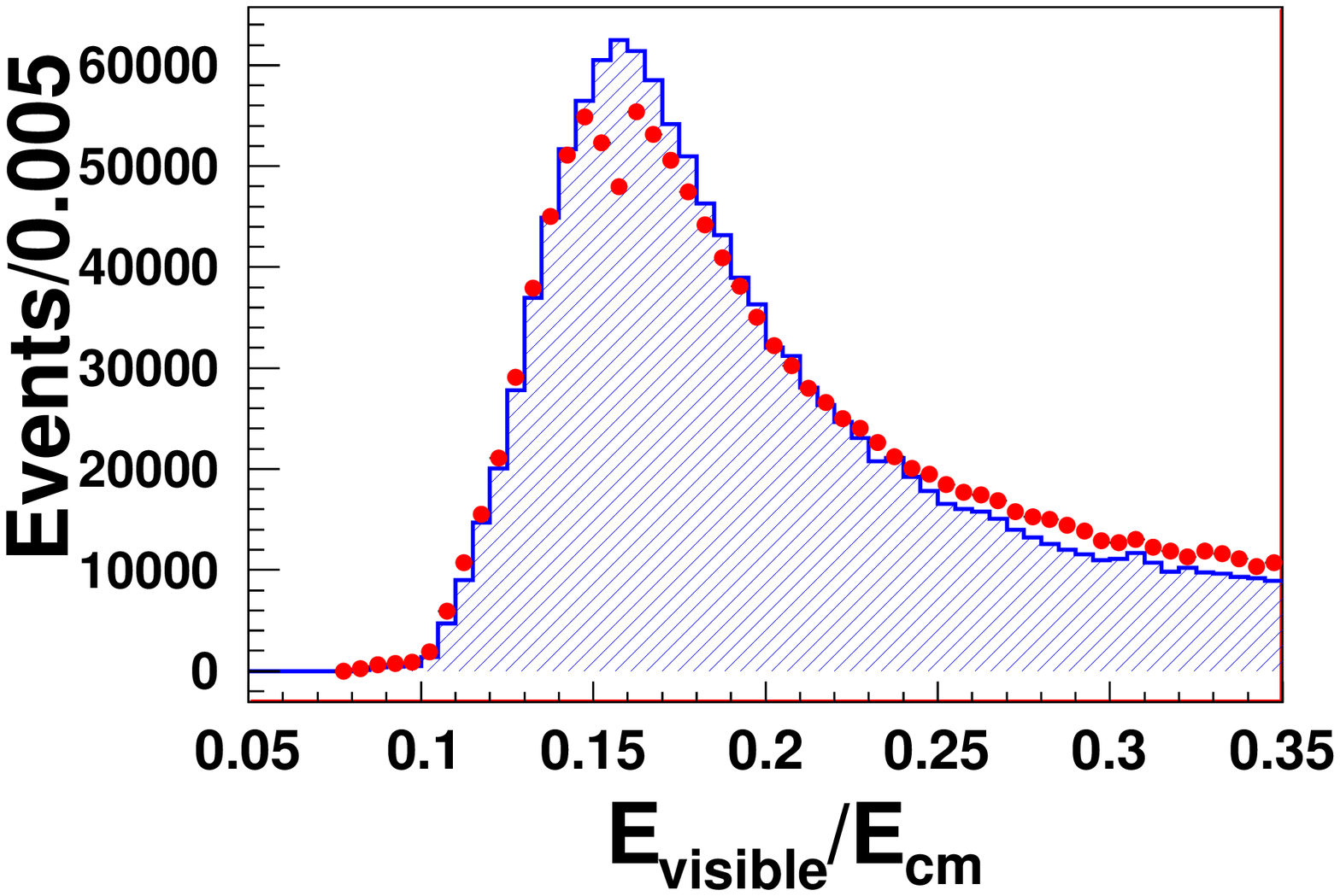}
\figcaption{Comparison of the $E_{\rm visible}/E_{\rm cm}$
distributions for the type-I (top) and type-II (bottom) LEB events
between the $\psp$ and scaled off-resonance data. The dots with error bars denote the former, and the shade histogram denotes the latter.}
\label{ebkg}
\ecl

Table~\ref{npsppart} shows the numbers of the observed hadronic
events for different charged-track multiplicity requirements of
the $\psp$ and ($N^{\rm obs}_{\psi(3686)}$) and off-resonance data ($N^{\rm obs}_{\rm off-resonance}$), as well as the remaining number of $\EE\ar\TT$ events ($N^\text{uncanceled}_{\TT}$)  estimated from MC simulation.
The corresponding detection efficiencies of $\psp\ar{\rm hadrons}$ are determined with $363.7\times 10^6$ $\psp$
inclusive MC events, and are listed in this table. The branching fraction of $\psp\ar {\rm hadrons}$ is included in the efficiency.
Figures~\ref{ncharge} show the comparisons for $\cos\theta$, $E_{\rm visible}/E_{\rm cm}$,
charged-track multiplicity, and photon multiplicity distributions
after background subtraction between data and MC simulation, a reasonable good agreement between data and MC simulation are observed.
\end{multicols}

\ruleup
\begin{table}[htbp]
\bcl \caption{Numbers of the observed hadronic events and the total numbers of $\psp$ events ($\times 10^6$),
the detection efficiencies of $\psp\ar {\rm hadrons}$ for different charged-track multiplicity requirements.}
\begin{tabular*}{110mm}{ccccccc}\\\hline\hline
Multiplicity &\multicolumn{2}{c}{$N_{\rm good}\geq
1$}&\multicolumn{2}{c}{$N_{\rm good}\geq
2$}&\multicolumn{2}{c}{$N_{\rm good}\geq 3$}\\\hline
Year
&2009&2012&2009&2012&2009&2012\\\hline
$N^{\rm obs}_{\psp}$ &107.72&343.51&103.72&329.04&82.28&259.98\\
$N^{\rm obs}_{\rm off-resonance}$ &2.23&1.325&2.01&1.245&0.74&0.400\\
 $N^{\rm uncanceled}_{\TT}$&0.036&0.57&0.034&0.54&0.013&0.21 \\
$\epsilon$(\%)&92.92&92.39&89.96&88.96&74.73&73.20\\\hline
$N_{\psp}$~~&107.2&341.7&107.2&340.5&106.6&343.6\\\hline\hline
\end{tabular*}
\label{npsppart} \ecl
\end{table}

\bcl
\includegraphics[width=6.0cm]{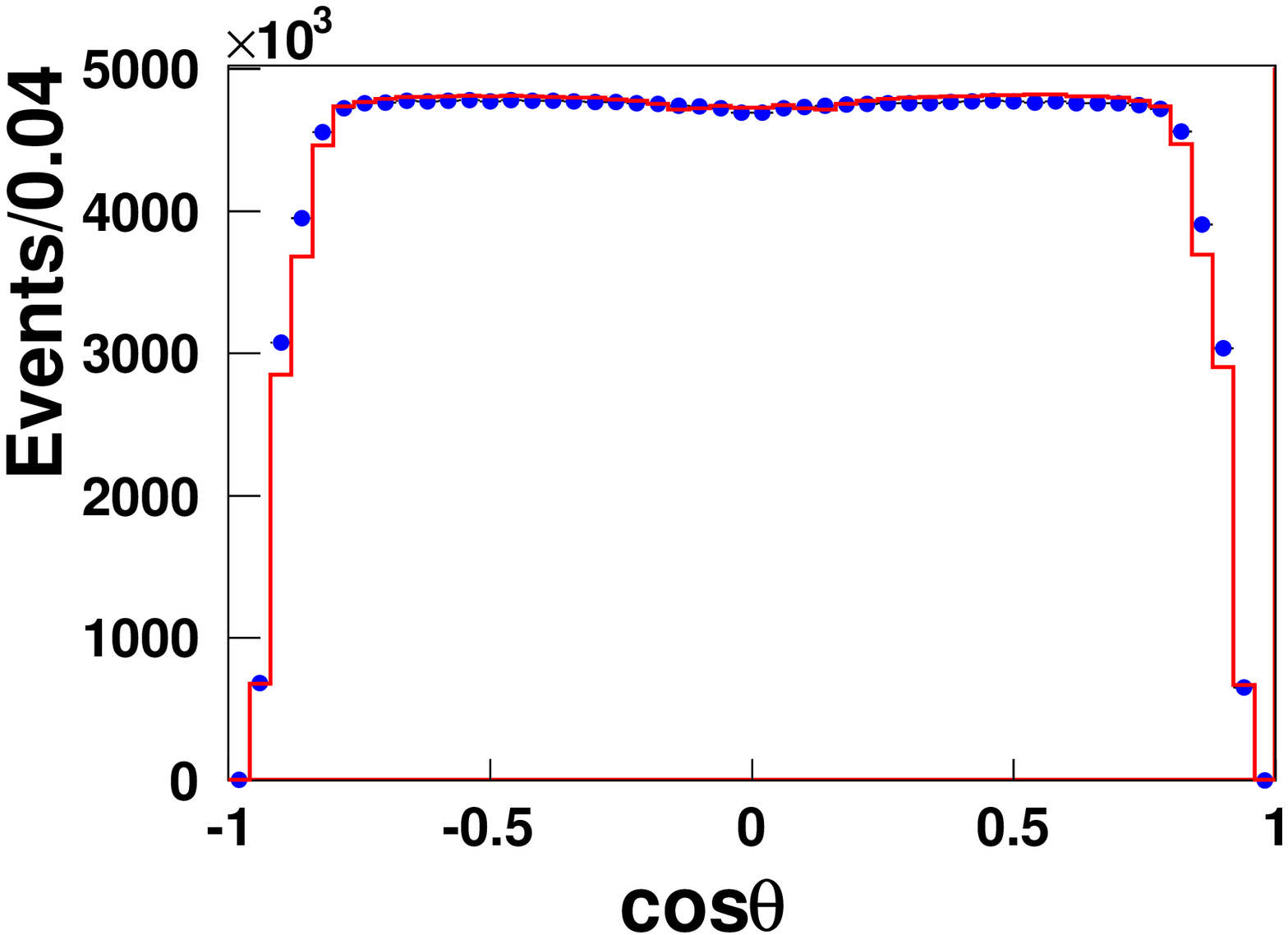}
\includegraphics[width=6.0cm]{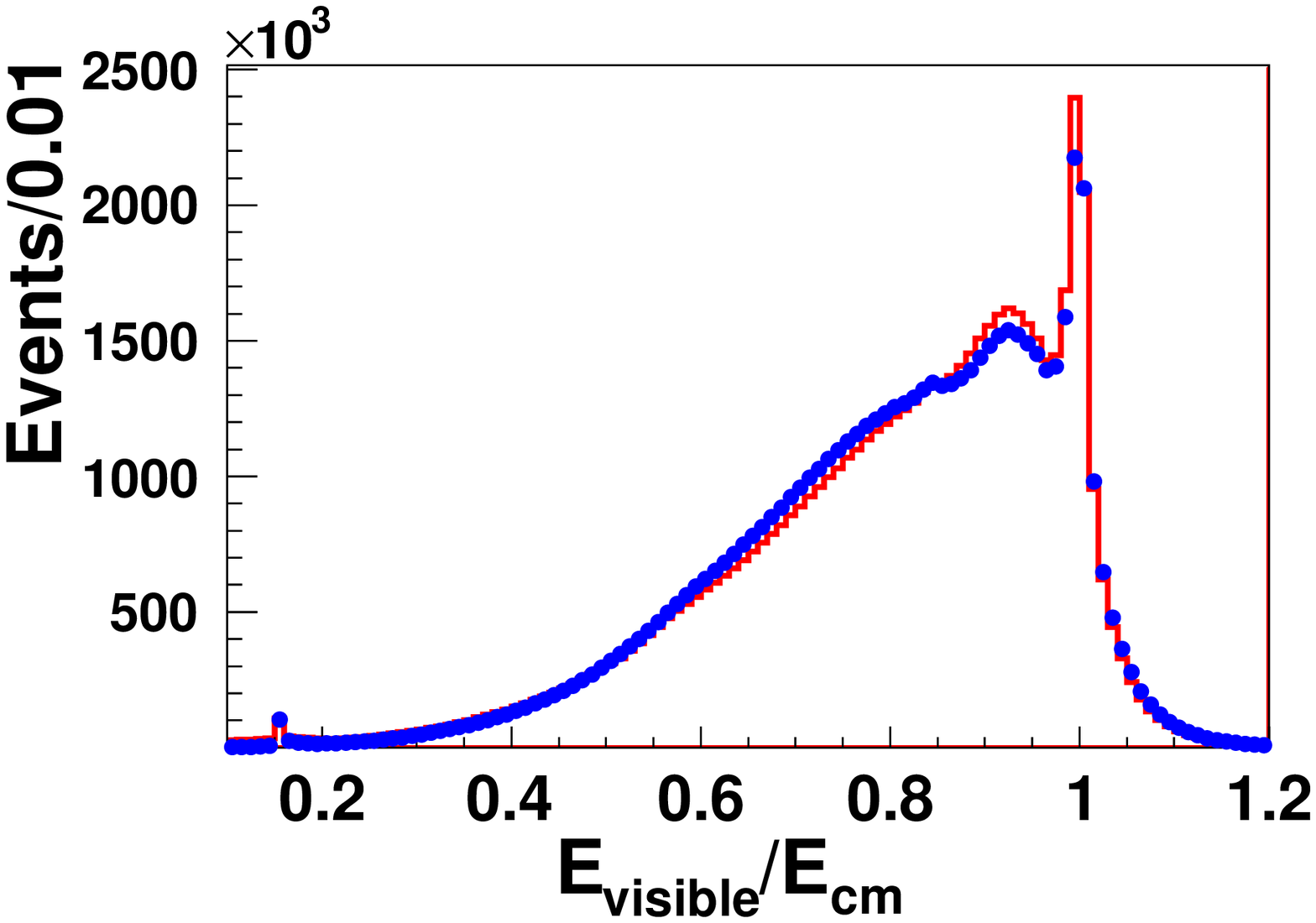}
\includegraphics[width=6.0cm]{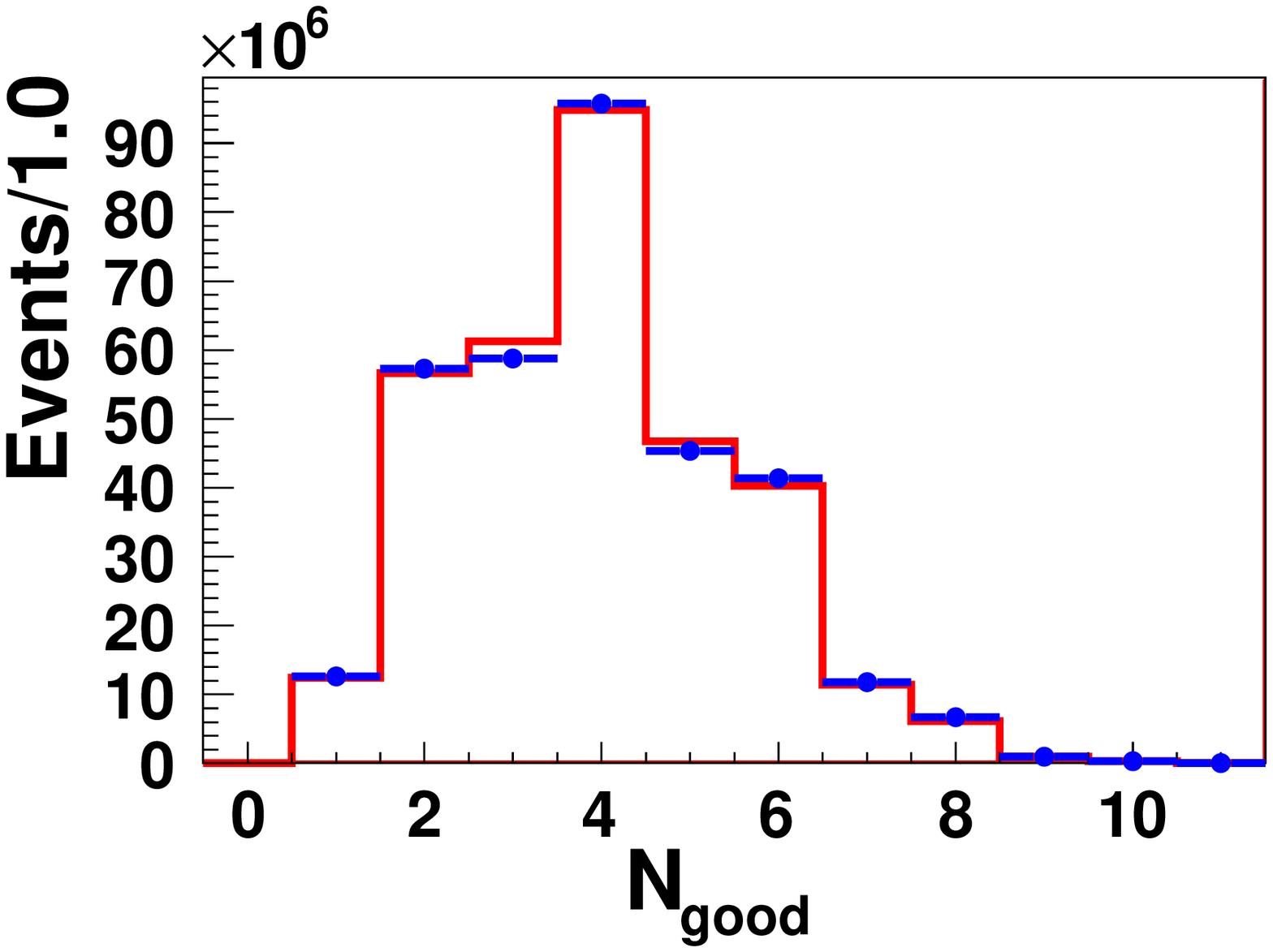}
\includegraphics[width=6.0cm]{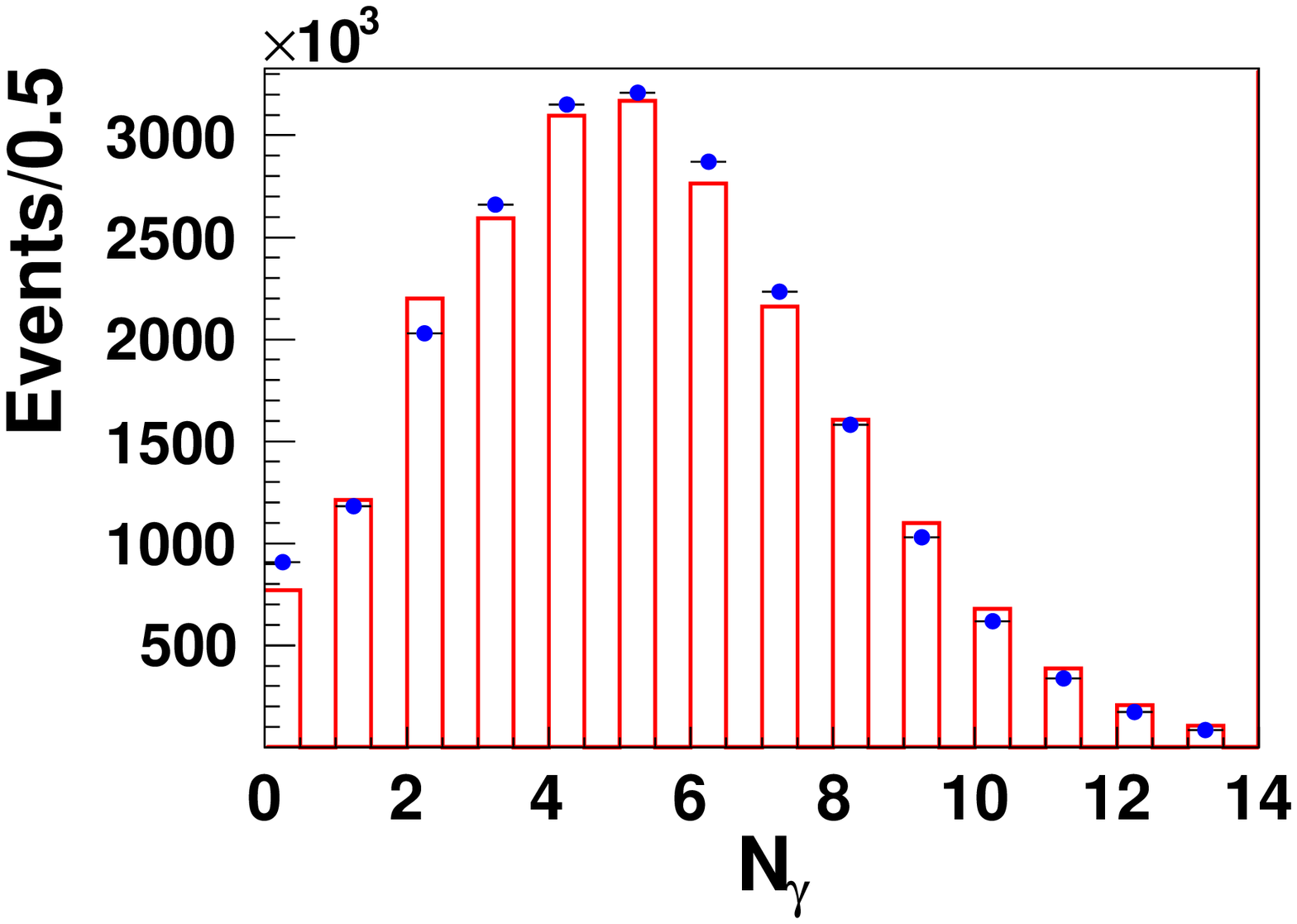}
\figcaption{\label{ncharge} The comparison of data/MC. (top-left) The $\cos\theta$
distribution. (top-right) The $E_{\rm visible}/E_{\rm cm}$
distribution. (bottom-left) The charged-track multiplicity
distribution. (bottom-right) The photon multiplicity distribution.}
\ecl
\begin{multicols}{2}
\ruleup

\section{Numerical results}
The total number of $\psp$ events, $N_{\psp}$, can be  calculated from
\beq\label{npsip}
N_{\psp}=\frac{N_{\rm peak}^{\rm obs}-f\cdot N_{\rm off-resonance}^{\rm obs}-N_{\TT}^{\rm uncanceled}}{\epsilon},
\eeq
With the numbers listed in Table~\ref{npsppart},  the numerical results for $\npsp$ with different charged-track multiplicity requirement are calculated and listed in Table I, too. We can see that there are slight differences between different multiplicity requirements due to the imperfect MC simulation on the charged tracks multiplicity. To obtain a more exact numerical result of $N_{\psp}$, an unfolding method is employed based on an efficiency matrix, whose matrix element, $\epsilon_{ij}$, represent the probabilities to observed $i$ charged tracks for an event with really $j$ charged tracks. The efficiency matrix is extracted from the inclusive MC samples. In practice, there are even numbers of charged tracks generated in an events due to the charge conservation,  while any number of charged tracks can be obtained due to the reconstruction efficiency and backgrounds. Therefore, the true charged track multiplicity of data sample is estimated from the observed multiplicity and the efficiency matrix
by minimizing a $\chi^2$ value, defined as
\beq
\chi^2 =\sum\limits^{10}_{i=1}\frac{(N_i^{\rm
obs}-\sum\limits_{j=0}^{10}\epsilon_{ij}\cdot N_j)^2}{N_i^{\rm
obs}},
\eeq
where the values $N_j~(j=0,~2,~4,\cdots)$ are the
true multiplicities of charged tracks in the data sample. They are the free parameters in the fit. For simplicity, the events with
ten or more tracks are considered in a single value, $N_{10}$. The
$N_{\psp}$ can be calculated by summing over all the obtained $N_{j}$. The results
are $107.0\times 10^6$ and $341.1\times 10^6$ for the 2009 and 2012
data samples, respectively.

\section{Systematic uncertainties}
The systematic uncertainties on the $\npsp$ measurement from different sources are described below and listed in Table~\ref{bkgxx}. The total systematic uncertainty is determined by the quadratic sum of all individual values.
\subsection{Polar angle}
The polar angle acceptance  for the charged tracks in the MDC is
$|\cos\theta|<$0.93. From Fig.~\ref{ncharge} (top-right), one finds
a slight difference between data and MC simulation at large polar
angles. As a check, we change the requirement on the polar angle to
be $|\cos\theta|<$0.8. The difference in $N_{\psp}$  is taken as the uncertainty due to the requirement on the polar angle.

\subsection{Tracking}
A small deviation (less than 1\%) on the tracking efficiency  between data and MC simulation is observed by various studies~\cite{track}. Assuming the average efficiency difference between data and MC simulation is 1\% per track, the effect can be studied by randomly removing every MC simulated tracks with 1\% probability. This results in a negligible difference in $N_{\psp}$, implying that $N_{\psp}$ is not sensitive to the tracking efficiency.
\subsection{Charged-track multiplicity}
The effect due to the simulation of the charged-track
multiplicity has been taken into account by the unfolding method described above. By comparing
the results between the direct calculation in Table~\ref{npsppart}
and the unfolding method including the  $N_{\rm good}\leq 1$ events, one
finds a difference of about 0.2\% on $N_{\psp}$ for both 2009 and 2012 data, which
is taken as the uncertainty associated with the charged-track
multiplicity.

\subsection{Momentum and opening angle}
For the type-II events, the requirements on momentum of charged tracks and opening angle between two charged tracks are applies to reject the sizable background from of Bhabha and dimuon events effectively.  When the requirement of charged track momentum is changed from $P<1.7$
GeV/$c$ to $P<1.55$ GeV/$c$, the resultant change on $N_{\psi(3686)}$ is negligible. When the requirement of opening angle between two charged tracks is changed from
$\theta<176^{\circ}$ to $\theta<160^{\circ}$, the change in $\npsp$
is negligible small for the 2009 data and is 0.04\% for the 2012
data, respectively. Figures~\ref{angle}  shows the comparisons of the distribution with background subtraction of the momenta and opening angles of the two charged tracks in the type-II events between the data and inclusive MC simulation.

\bcl
\includegraphics[width=6.0cm]{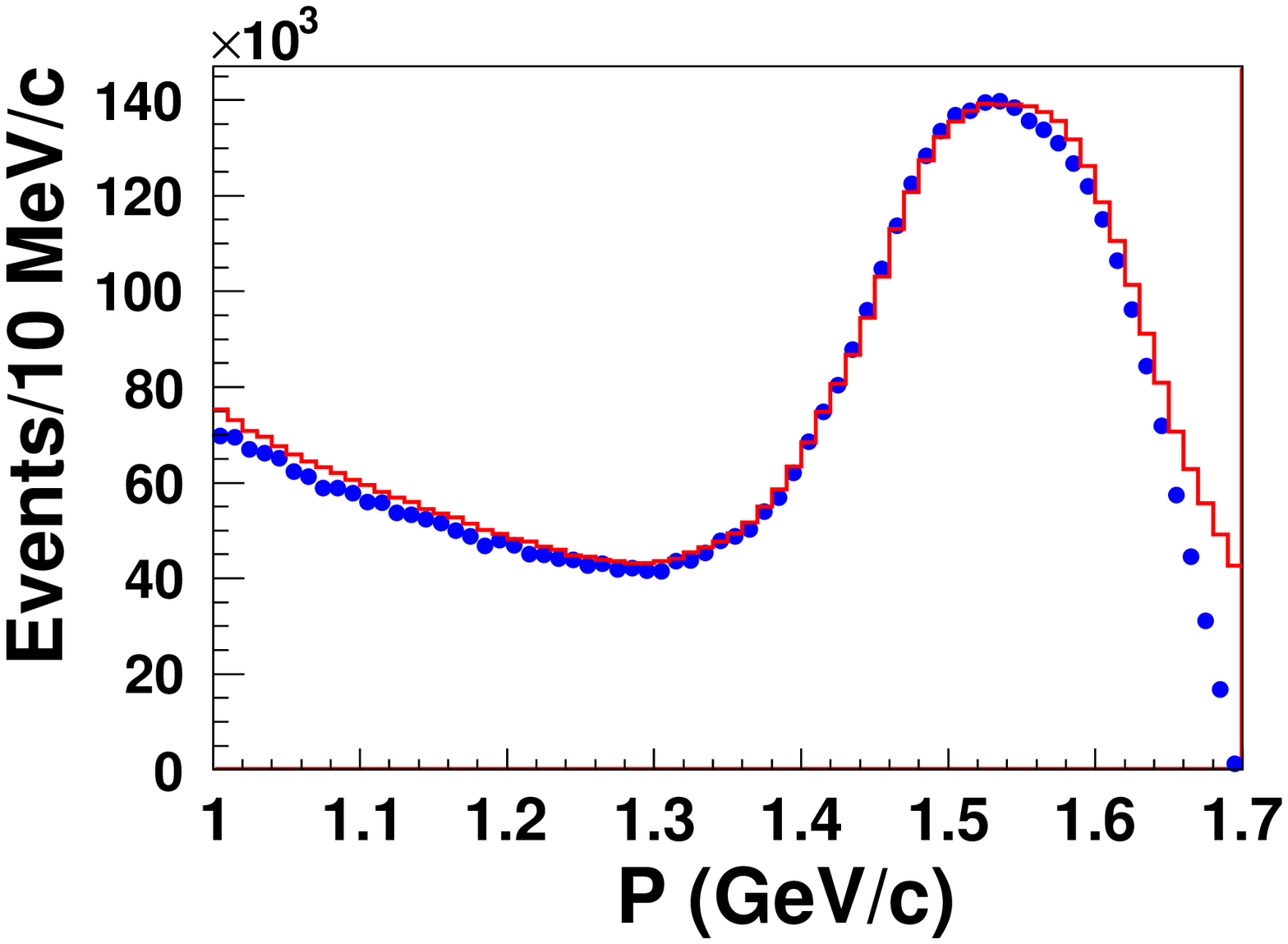}
\includegraphics[width=6.0cm]{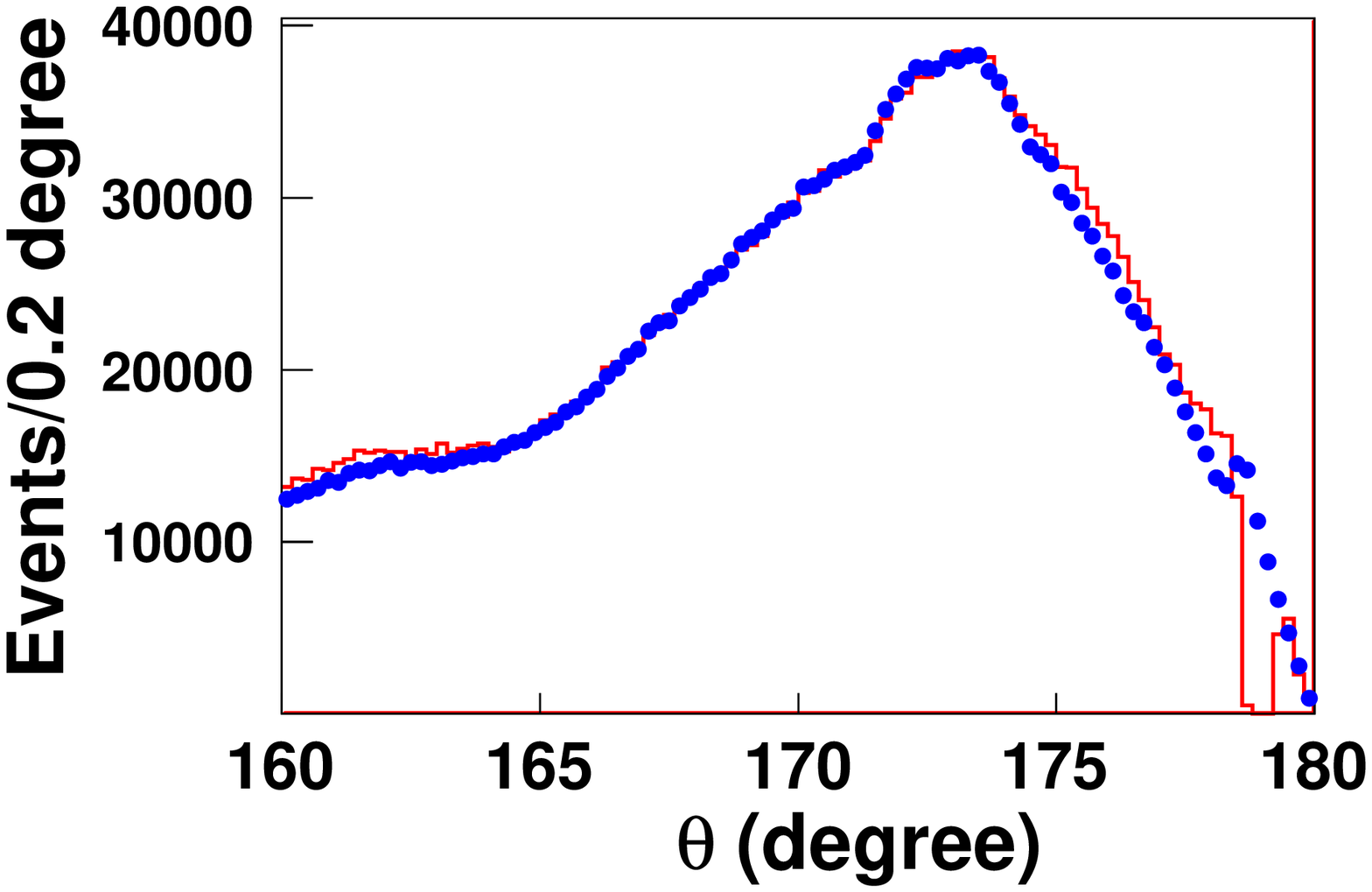}
\figcaption{\label{angle}Distributions of charged track momentum (top)
and opening angle between the two charged tracks (bottom) for the type-II
events.}
\ecl

\subsection{LEB contamination}
$\npsp$ is insensitive to the visible energy requirement. The uncertainty associated with the requirement $E_{rm visble}/E_{rm cm}>0.4$ is estimated by comparing the results with or without this requirement, the difference on $N_{\psi(3686)}$  is assigned to be the corresponding uncertainty.

\subsection{Determination of $N^{\rm obs}$}
As mentioned as in Sec. 3, two methods are used to obtain $N^{\rm obs}$. The nominal method counts the numbers of events in the signal region and subtracts the number of background estimated in the sideband regions. The alternative method is performed by fitting the $\bar{V}_Z$ distribution.
The resultant difference on $N^{\rm obs}$ between these two methods is taken as the uncertainty
in the determination of $N^{\rm obs}$.

\subsection{Vertex limit}
We repeat the analysis by changing the requirement $V_r<1\;\text{cm}$ to $V_r<2\;\text{cm}$, the change on $N_{\psi(3686)}$ is small and is taken as the systematic uncertainty. Similarly, we repeat the analysis by changing the requirement $|\bar{V}_Z|<10\;\text{cm}$ to $|\bar{V}_Z|<20\;\text{cm}$, and find a negligible change on $\npsp$.

\subsection{Scaling factor}
The scaling factor ($f$) for the background subtraction depends on the luminosity of data samples.
In the nominal analysis, the luminosity is estimated with the $\EE\ar\GG$ events.  Alternative measurement on the luminosity is performed with the large angle Bhabha events, and the scaling factor as well as the $N_{\psi(3686)}$ are recalculated.
The resultant difference in $N_{\psi(3686)}$ is found to be negligible, and the corresponding uncertainty is not considered.

\subsection {Choice of sideband region}
In the nominal analysis, we take $|\bar{V}_Z|<4;\text{cm}$ as the signal region and
$6<|\bar{V}_Z|<10\;\text{cm}$ as the sideband region. A alternative analysis is repeated by shifting the sideband region outward by 1~cm, which is about 1$\sigma$ of the $\bar{V}_Z$ resolution. The resulting difference in $N_{\psi(3686)}$ is taken as the systematic uncertainty.

\subsection {\boldmath $\pi^0$ mass requirement}
The $\pi^0$ mass requirement is only applied for the type-I events.
There is a slight change in $\npsp$ when the mass window requirement
is changed from $|M_{\GG}-M_{\pi^0}|<0.015\;\text{GeV}/c^2$ to
$|M_{\GG}-M_{\pi^0}|<0.025\;\text{GeV}/c^2$. This difference is taken as
the uncertainty due to the $\pi^0$ mass requirement.

\subsection{The missing 0-prong hadronic events}
A detailed topological analysis is performed for the events with $N_{\rm
good}=0$ in the inclusive MC sample. Most of these events
come from the well-known decay channels, such as $\psp\ar
X+\jpsi$~(where X denotes $\eta, \pi^0,\pi^0\pi^0, \gamma\gamma$~$etc.$)
and $\psp\ar\EE$, $\MM$. The fraction of these 0-prong events
in the inclusive MC sample is $\sim$2.0\%, of which the pure neutral
channels contribute about 1.0\%. As shown in Fig.~\ref{ncharge}, the MC simulation models data well.  Therefore, we
investigate the pure neutral hadronic events, which are selected according
to the following scheme. With the same charged track and shower
selection criteria as above, we require $N_{\rm good}=0$ and
$N_{\gamma}>3$. The latter requirement is used to suppress
$\EE\ar\GG$ and beam-associated background events. The same
selection criteria are imposed on the off-resonance data and inclusive MC events.
Figure~\ref{0pr} shows the distributions of the total energies in
 the EMC, $E_\text{EMC}$, for the different data sets and inclusive MC sample. The peaking events around the center-of-mass energy are taken as the pure neutral hadronic candidates. As shown in Fig.~\ref{0pr}, the number of signal events is extracted by a fit on the $E_\text{EMC}$ distribution.  In this fit, the signal is described by a Crystal Ball function, the QED background in $\psi(3686)$ data is described by the shape of
 off-resonance data (off-resonance data at $ \sqrt{s}= 3.65\;\text{GeV}$ or
 $\tau$-scan data) after scaling for luminosity, and the other backgrounds are described by a
 polynomial function. For 2012 data, the difference in the number of pure neutral hadronic events
 between the data and the inclusive MC simulation sample is 11\% if the $\tau$-scan data sample is taken as the off-resonance data to estimate the background function, as shown in Fig.~\ref{0pr} (top).
 However, this difference changes to 18\% if we use the
 off-resonance data at $\sqrt{s}=3.65\;\text{GeV}$ for the background
 function, as shown in Fig.~\ref{0pr} (middle). The larger difference is used to estimate the uncertainty conservatively. Since the fraction of the pure neutral hadronic events is about 1.0\% of the total selected candidates, the uncertainty due to the missing 0-prong events should be less than $18\%\times1\% = 0.18\%$ for the 2012 data. The same method is applied to the 2009 data samples, and the uncertainty is 0.25\%, which is found to be somewhat larger than the previous analysis~\cite{psp09}.

\bcl
\includegraphics[width=5cm]{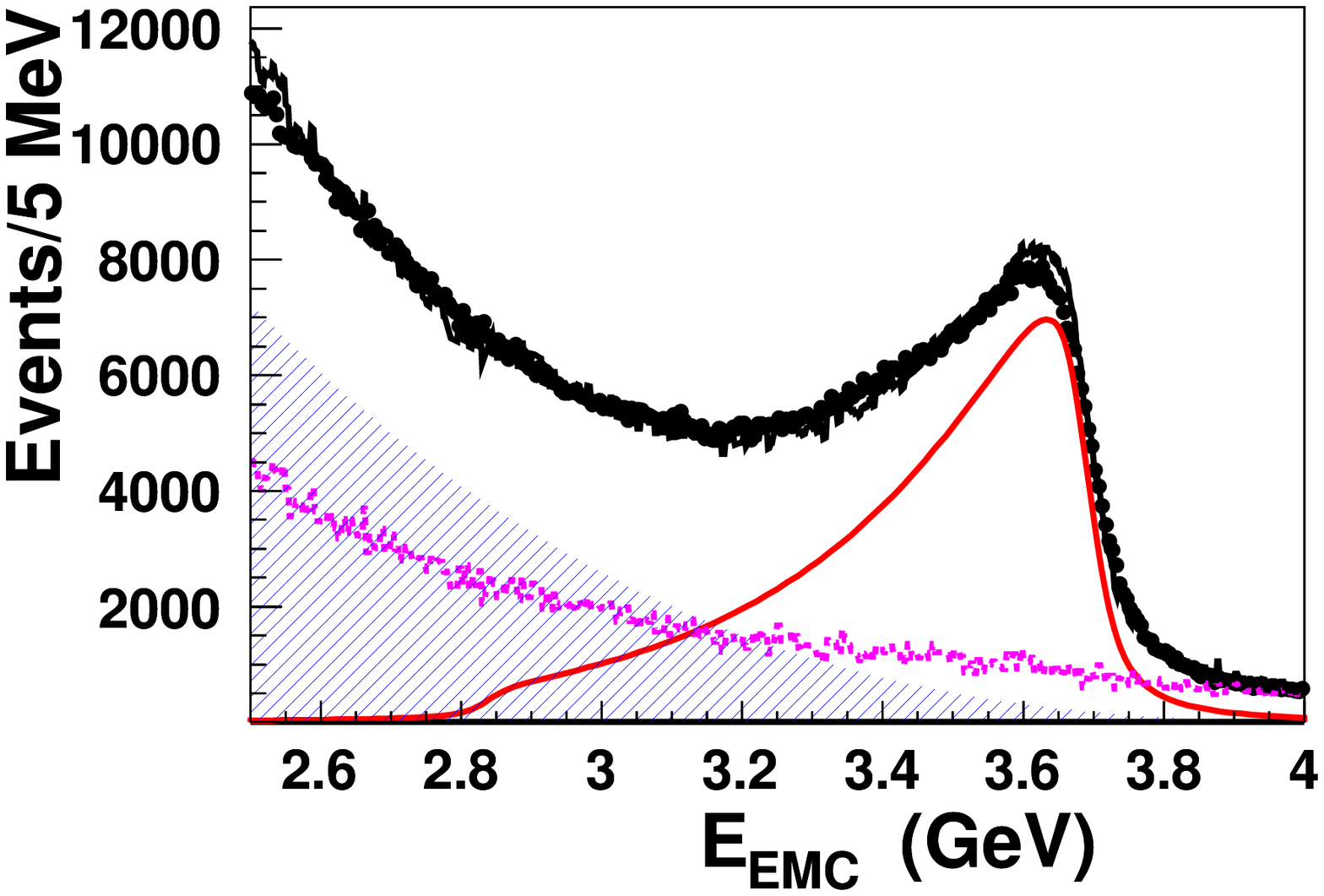}
\includegraphics[width=5cm]{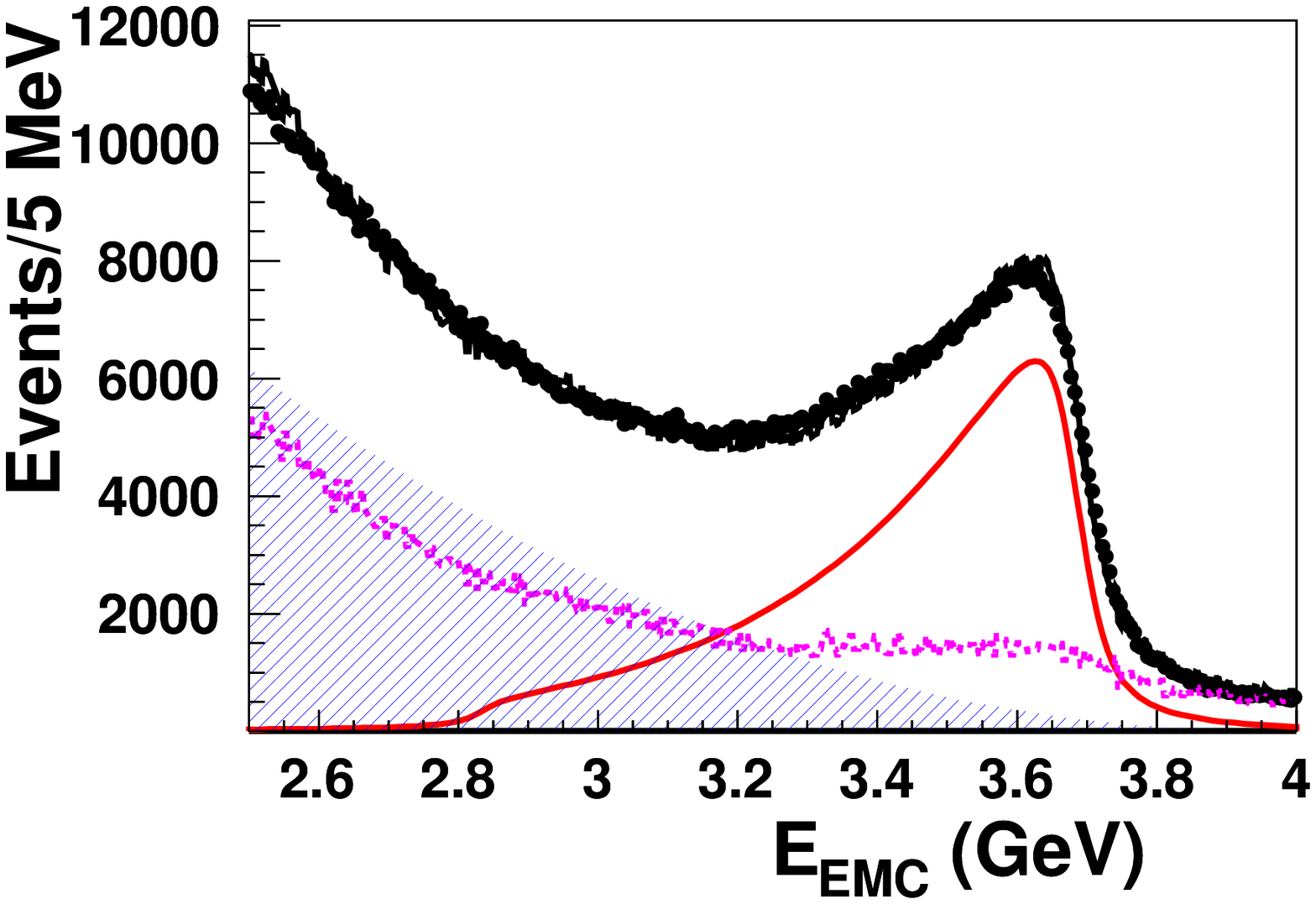}
\includegraphics[width=5cm]{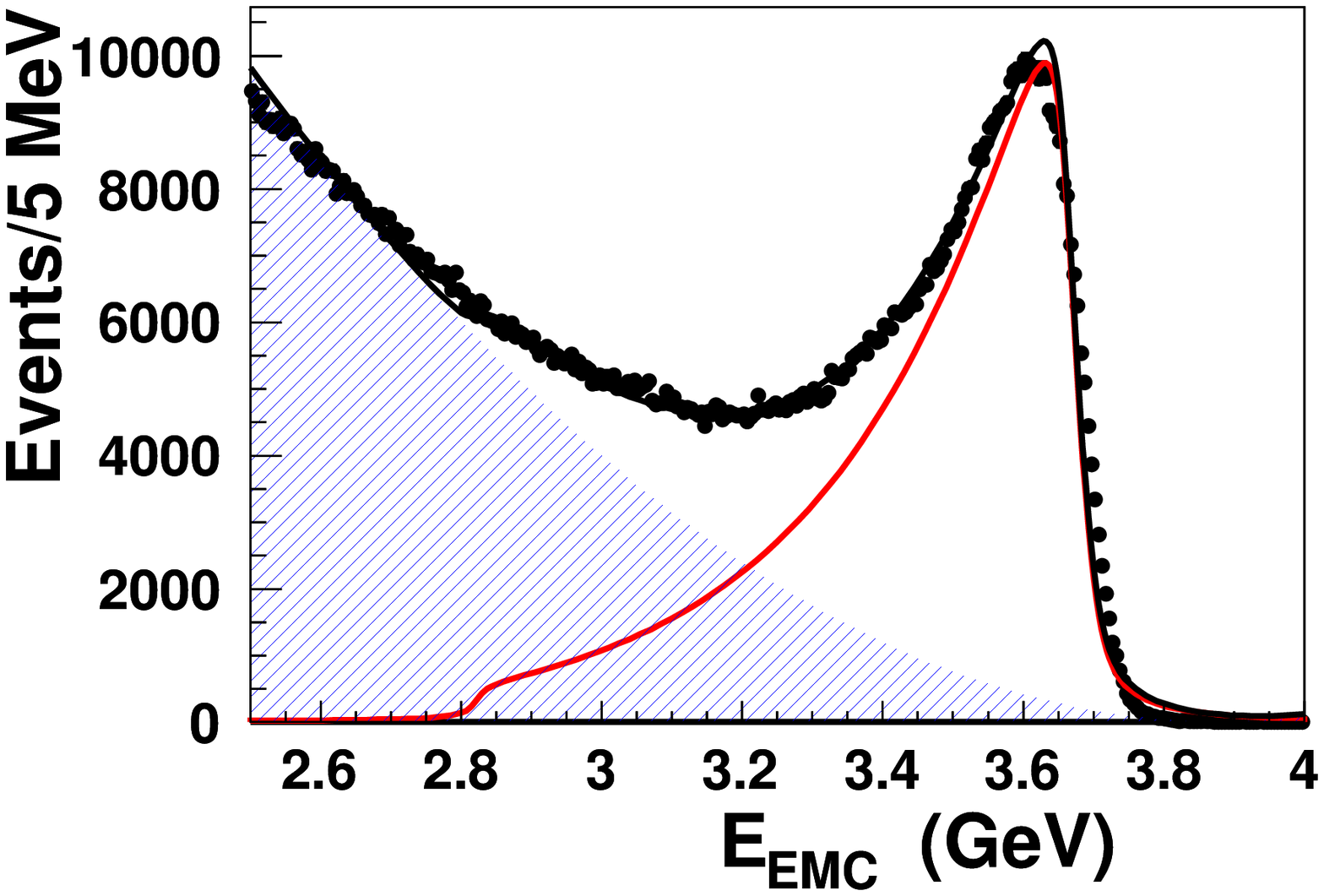}
\figcaption{\label{0pr}Distributions of the total energies in
the EMC for the $N_{\rm good}=0$ events
for the $\psp$ data with QED
background approximated by the $\tau$-scan data (top),
the data taken at $\sqrt{s}=$3.65 GeV (middle), and
the inclusive $\psp$ MC sample (bottom).
The dot-dashed lines denote the signal shapes of
neutral $\psp$ decays and the shaded regions are the background
shapes from $\psp$ decays. The dashed lines denote the background
shapes from QED processes.}
\ecl

\subsection{MC modeling}
The uncertainty due to the MC simulation of inclusive $\psp$ decays
arises from sources such as the input of branching ratios, the angular
distributions of the known and unknown decay modes, $etc$. Actually, the possible related uncertainty have been covered by those from the charged-track multiplicity, missing of 0-prong events $etc$, whose uncertainties have been studied carefully. Thus, no further uncertainty is assigned for the MC modeling.

\subsection{Trigger}
Based on the 2009 data, we have studied and found that the trigger efficiency
for the $N_\text{good}\geq2$ (type-II and type-III) events is close to 100.0\%, while it is 98.7\% for the type-I events~\cite{nik}. Since the fraction of type-I events is only about 3\% of the total selected events, the uncertainty caused by the trigger is negligible for 2009 data. As shown in Table~\ref{npsppart}, the fraction of type-I events in 2012 data is the same as that in 2009 data. Furthermore, an additional neutral trigger channel was added during 2012 data taking. Therefore, the trigger efficiency for the 2012 data is expected to be higher for type-I events than that for 2009 data, and the uncertainty associated with the trigger can be neglected.

\subsection{\boldmath $B(\psp\ar {\rm hadrons})$}
The uncertainty of the branching ratio for $\psp\ar {\rm hadrons}$
is small, 0.13\% quoted from PDG~\cite{PDG}, and is taken as the uncertainty.

\end{multicols}

\begin{table}
\bcl \caption{Summary of systematic uncertainty (\%).} \doublerulesep 2pt
\begin{tabular*}{100mm}{l@{\extracolsep{\fill}}lcc}
\toprule Source&2009&2012\\\hline
Polar angle& 0.27&0.31\\
Tracking & negligible&negligible\\
Charged-track multiplicity&0.20&0.19\\
Momentum and opening angle&negligible&0.04\\
LEB contamination&negligible&0.09\\
$N^{\rm obs}$ determination&0.27&0.30\\
Vertex limit &0.32&0.21\\
Scaling factor ($f$) & negligible&negligible\\
Choice of sideband region&0.32&0.26\\
$\pi^0$ mass requirement &0.09&0.05\\
0-prong~ events&0.25&0.18\\
Trigger& negligible&negligible\\
MC modeling& negligible&negligible\\
$B(\psp\ar {\rm hadrons})$&0.13&0.13\\\hline
Total&0.70&0.63\\\hline\hline
\end{tabular*}
\label{bkgxx}
\ecl
\end{table}

\begin{multicols}{2}
\ruleup
\section{Summary}
The number of $\psp$ events taken by BESIII in 2012 is measured to be
$(341.1\pm 2.1)\times 10^6$ with the inclusive hadronic events, where the uncertainty is dominated by systematics, and the statistical uncertainty is negligible. The
number of $\psp$ events taken in 2009 is also updated to be $(107.0\pm 0.8)\times 10^6$,
The slight difference, but consistent within the uncertainty, in the mean of number of events with respect to the previous measurement and the much improved precision are due to the refined offline software, MC tuning, and
the method of $N_{\psp}$ determination. Adding them linearly
yields the total number of $\psp$ events for the two runs data taking to be
$(448.1\pm 2.9)\times 10^6$. This work provides a basic and important
parameter for the studies of the decays of the $\psi(3686)$ and its
daughters.

\section{Acknowledgment}
The BESIII collaboration thanks the staff of BEPCII and the IHEP
computing center for their strong support. This work is supported in
part by National Key Basic Research Program of China under Contract
No. 2015CB856700; National Natural Science Foundation of China
(NSFC) under Contracts Nos. 11235011, 11322544, 11335008,
11425524, 11475207; the Chinese Academy of Sciences (CAS) Large-Scale
Scientific Facility Program; the CAS Center for Excellence in
Particle Physics (CCEPP); the Collaborative Innovation Center for
Particles and Interactions (CICPI); Joint Large-Scale Scientific
Facility Funds of the NSFC and CAS under Contracts Nos. U1232201,
U1332201, U1532257, U1532258; CAS under Contracts Nos. KJCX2-YW-N29, KJCX2-YW-N45; 100
Talents Program of CAS; National 1000 Talents Program of China;
INPAC and Shanghai Key Laboratory for Particle Physics and
Cosmology;  German Research Foundation DFG under Contracts Nos. Collaborative Research Center CRC 1044, FOR 2359; Istituto Nazionale di Fisica Nucleare, Italy; Koninklijke
Nederlandse Akademie van Wetenschappen (KNAW) under Contract No.
530-4CDP03; Ministry of Development of Turkey under Contract No.
DPT2006K-120470; The Swedish Resarch Council; U. S. Department of
Energy under Contracts Nos. DE-FG02-05ER41374, DE-SC-0010504,
DE-SC-0012069; U.S. National Science Foundation; University of
Groningen (RuG) and the Helmholtzzentrum fuer Schwerionenforschung
GmbH (GSI), Darmstadt; WCU Program of National Research Foundation
of Korea under Contract No. R32-2008-000-10155-0.

\end{multicols}

\end{document}